\documentclass[12pt,cites]{article}
\usepackage{latexsym,graphicx,amssymb,amsmath}
\usepackage{appendix}
\usepackage{cite}
\title{Electron Transfer Reaction Through an Adsorbed Layer}

\author{A. V. B. Cruz$^a$, A. K. Mishra$^a$ and 
W. Schmickler$^b$\\
$^a$ Institute of Mathematical Sciences\\CIT Campus, Chennai, 600113 India\\
$^b$ Institute of Theoretical Chemistry, Ulm University\\
D89069 Ulm, Germany}
\begin{document}
\maketitle
\vskip 2.0cm

\baselineskip=20pt

\begin{abstract}
 We consider electron transfer from a redox to an electrode through and adsorbed intermediate. The formalism is developed to cover all regimes of coverage factor, from lone adsorbate to monolayer regime. The randomness in the distribution of adsorbates is handled using coherent potential approximation. We give current-overpotential profile for all coverage regimes. We explictly analyse the low and high coverage regimes by supplementing with DOS profile for adsorbate in both weakly coupled and strongly coupled sector. The prominence of bonding and anti-bonding states in the strongly coupled adsorbates at low coverage gives rise to saddle point behaviour in current-overpotential profile. We were able to recover the marcus inverted region at low coverage and the traditional direct electron transfer behaviour at high coverage. 
\end{abstract}

\section {Introduction}

\hspace{0.2in}

A proper understanding of electron transfer reaction through an adsorbate intermediate 
constitutes the first step towards modelling the charge transfer across a chemically modified 
electrode \cite{durst,Metzger01,Salomon01}, through a molecular wire \cite{Jortner01,Muller01}, or the phenomenon of the molecular electronics 
\cite{Reed01,Roth,Joachim,Remacle01,Lindsay01}. In fact the indirect heterogeneous electron transfer is a recurring feature in all 
these processes.

\vspace{0.2in}

In the present communication, we consider the kinetics of an adsorbate mediated electron
transfer reaction. The adsorbate is taken to be a metal ion.  The reactant is supposed to couple with the adsorbate orbital alone; the 
direct coupling between the reactant and Bloch states in the metal electrode is neglected. 
In the present study, the adsorbate coverage factor $\theta$ is allowed to take any
arbitrary value in the range $(0,1)$. Thus starting from a single adsorbate case, corresponding to $\theta \longrightarrow 0$ limit,
the formalism remains valid all the way up to a monolayer regime   $    (\theta = 1)$  .
For metallic adsorbates, the adsorbate orbitals remain spatially localized
in the low coverage regime. But in the monolayer regime, one obtains extended electron states 
in the adlayer. These states now form a two-dimensional band \cite{mishra1,ramkishore01}. The localized adsorbate states
interact strongly with the solvent polarization modes. On the other hand, the interaction of 
extended electron states with polarization modes are much weaker, and as a first approximation,
it can be neglected \cite{mishra2} . 
This progressive desolvation of adspecies, when the coverage is varied from 
zero to one, changes adsorbate orbital energy by a few electron volts and 
hence must leave very significant effects on the electron kinetics. 
In addition, in the monolayer regime, the metallic adlayer
acts as the electrode surface. As a consequence, the adsorbate mediated electron transfer
exhibits the characteristics of a direct heterogeneous reaction. 
We investigate how the desolvation and metallization of the adsorbate layer  
influences the charge transfer kinetics.

\vspace{0.2in}

The crucial difference between the heterogeneous electron transfer reaction through an 
adsorbate and a direct electron transfer to an electrode arises due to (i) a possible
change in the electronic coupling term between the participating orbitals, and (ii)
modification in the relevant density of state. The electronic coupling strength can 
change due to the particular symmetry configuration of the orbitals.  Besides, the 
equilibrium distance between the reactants may vary in both the situations [  ]. Next,
the density of states (dos) of a metallic electrode is usually broad, and a slowly varying
function of energy [  ]. This feature enables one to replace the energy dependent
dos by its value at the Fermi level. As the coupling between the band states in the 
electrode and solvent polarization modes are usually neglected, the electrode dos
does not exhibit any temperature dependence. The adsorbates, on the other hand
have a narrower dos, which depends on temperature.
This follows from the solvent induced broadening of the adsorbate orbital. When 
this broadening mechanism is absent, adsorbate dos ceases to be temperature dependent.
In addition to the solvent induced broadening, the adsorbate dos acquires an 
additional temperature independent  width due to the hybridization of 
its orbital with the Bloch states in the electrode. The location of adsorbate dos 
vis a vis the dos of redox couple play an important role in determining the charge 
transfer kinetics. In fact catalytic effect can be observed when strong overlap  
occurs between these two density of states.

\vspace{0.2in}

The adsorbates  exhibit different structural arrangements at different coverage.
Even at a fixed coverage, more than one kind of distribution pattern can be
observed in the adlayer \cite{Neugebauer01,Neugebauer02,Over01}. Modelling each configuration separately poses
a difficult task. Therefore we consider a random distribution of the adsorbates 
in a two dimensional layer. Subsequently, an `effective-medium' description is 
used for the adlayer. This procedure captures the essential features
of the adlayer in an average sense.

\vspace{0.2in}

{\small The plan of the paper is as follows: In section 2, we present the model Hamiltonian and
the expression for anodic and cathodic current in terms of system parameters, whose detailed calculations as 
shown in appendix. In section 3, the results of 
numerical analysis is presented along with the various DOS for different regimes and also the profile of current 
at different coverages are considered along with explanation for the observed behaviour. In section 4, we summarize 
our results and an overview of the whole work is given}

\section {System Hamiltonian and Current}

\hspace{0.2in}

An adsorbate has  strong electronic coupling with  the substrate band states  
as well as it has electronic overlap with  the neighbouring adspecies. The latter coupling leads to a 
two-dimensional band formation in the adlayer at higher coverage.  The solvent-adsorbate
interaction and surface plasmon-adsorbate interaction, both modelled
within the harmonic approximation, are responsible for the  solvation and 
image energy for the adsorbate, respectively. Similar interactions are  present
for the redox-couple, which is supposed to interact weakly with the adsorbate
orbital. Taking into account various system components and interactions among them,
an effective Hamiltonian for an adsorbate mediated electron transfer reaction can be 
written as

\begin{equation}
H  =   \sum_\sigma \bar { \epsilon}_{r}\{b_{\nu} + b^\dagger_\nu \} n_{r\sigma} +
 \sum_{\sigma} \{v_{ar} c_{a\sigma}^{\dagger} c_{r\sigma} + h.c.\}
- \sum_{\nu = 1,2} {\lambda_{rc\nu}} (b_\nu + b^{\dagger}_{\nu}) + H_o 
\end{equation}
The redox species is coupled to an adsorbate located at a site $ i = a$ in the adlayer.
$H_o$ is the Hamiltonian for the `electrode - adsorbate - solvent' subsystem

\begin{equation}  \begin{array}{lcl}
H_o
 & = & \sum_{k, \sigma} \epsilon_k n_{k\sigma} + 
   \sum_{i,\sigma} \hat {\epsilon}_{i\sigma}(\{b_{\nu} + b_{\nu}^{\dagger}\})
 n_{i\sigma} +
  \sum^4_{\nu= 1 } \omega_\nu b^{\dagger}_\nu b_\nu \\ 
\\
& & + \sum_{k,i,\sigma} \{v_{ik} c_{i\sigma}^{\dagger} c_{k\sigma} + h.c.\} + \sum_{i \ne j
, \sigma }v_{ij} c_{i\sigma}^{\dagger} c_{j\sigma} \\ 
\\
& & - \sum_{i,\nu =1,2}  \lambda_{ic\nu} (b_\nu + b^{\dagger}_\nu) 
\\
\end{array}
\end{equation}
$\{i\}$ specify sites in the adlayer. $k$ and $r$ label electrode and reactant electronic states.
$\sigma$ is the spin index and $\epsilon$ is the energy value.
$n, c^{\dagger}$ and $c$ respectively denote number, creation 
and annihilation operators for electrons.  $\nu = 1,2,3,4$ label oscillator modes corresponding to 
the orientational, vibrational, electronic solvent polarization and surface plasmons,
respectively, and $\omega_{\nu}$ are the associated frequencies.
$ b, b^{\dagger}$ are the annihilation and creation operators for the boson modes.
$v$  represent the coupling term between the electronic states. $\lambda $ signify
the strength of adsorbate and reactant coupling with the boson modes. Subscripts $o$ and $c$
respectively refer to the reactant and adsorbate core.

\begin{equation}
{\bar { \epsilon}_{r} \{ b_\nu + b^\dagger_\nu \}}
 = \epsilon^o_{r} +  \sum_{\nu=}   \lambda_{r\nu} (b_\nu + b^{\dagger}_\nu) 
\end{equation}

\begin{equation}
\hat{\epsilon}_{i \sigma} \equiv \hat{\epsilon}_{a \sigma} + \sum_{\nu} \lambda_{a \nu} (b_{\nu} + b^{\dagger} _{\nu}) 
\label{esite}
\end{equation}
$\epsilon^o_r $ and $\epsilon^o_a$ are the reactant and adsorbate orbital energies in the gas phase
The expression (4) gives the energy of adsorbate at site $i$ when it is occupied. 
In case  no adsorbate occupies the site $i$, the expectation value
\begin{equation}
 <\hat \epsilon_{i\sigma}> \longrightarrow \infty 
\end{equation}
which ensures no charge transfer through an unoccupied site. While evaluating the 
shift in adsorbate orbital due to its coupling to boson, the boson mediated interaction 
between different sites are neglected. Next, since the adsorption of a 
single type of  species is considered, we replace $\lambda_{i\nu}$ by $\lambda_{a\nu}$ and
$\lambda_{ic\nu}$ by $\lambda_{c\nu}$. 
The randomness associated with the site energy can be handled using the 
coherent potential approximation [   ]. 

\vspace{0.2in}

Treating  the magnitude of $v_{ar}$ to be a small quantity, the anodic current contribution
is obtained within the linear response formalism as $(cf. Appendix A)$

\begin{equation}
I_a \, = \, 2\, e \theta |v_{ar}|^2 \sqrt{  \pi } \hbar^{-1}\, \int^{\infty}_{- \infty} 
sgn(X_2(\epsilon, \theta)) \, (1 - f (\epsilon)) \, \, \rho_a^{\bf {an}} (\epsilon) \, \rho_r^{\bf {an}}(\epsilon)
d\epsilon     
\label{anodeI}
\end{equation}

where $f(\epsilon) =  (1 + exp(-\beta \epsilon))^{-1}$    is the Fermi distribution function. 
Here zero of the energy  scale is set to be at $\epsilon_f$ for a direct electrochemical electron
transfer reaction.  $\rho_a^{\bf {an}} (\epsilon) $ and  
and $\rho_r^{\bf {an}}(\epsilon)$ are the adsorbate and the reactant density of states. 

\begin{equation}
\rho_a^{\bf {an }}(\epsilon) \ = \ \frac{1}{2\sqrt{\pi P}} Re(w(z)) 
\end{equation}

\begin{equation}
w(z) \ = \ e^{-z^2} erfc(-iz) 
\end{equation}

\begin{equation}
P \ = \ \frac{ \displaystyle ( 4 E^r_a E^r_r - (E^r_{ar})^2)} {\displaystyle  4 \beta  E^r_r}
\end{equation}

\begin{equation}
Z \ = \ (-Q^{\bf {an}} + i|X_2(\epsilon,\theta)|)/(2\sqrt{P})\,, 
\end{equation}

\begin{equation}
Q^{\bf {an}} \ = \  X_1(\epsilon,\theta)- \epsilon^{0} _{a\sigma} + \sum_{\nu} \frac{\lambda_{a \nu} \bar{\lambda}_{\nu}}{\omega_{\nu}}   - \frac{\displaystyle (\epsilon - \epsilon^{0} _r + \sum_{\nu} \frac{\lambda_{r \nu} \bar{\lambda}_{\nu}}{\omega_{\nu}}) E^r_{ar} }
{\displaystyle 2 E^r_r}
\end{equation}
 $Re( w(z))$  denotes the real part of [w(z)].   $X_1$ and $X_2$ is obtained through the relation (cf. Appendix)

\begin{equation}
  X_1(\epsilon, \theta) + i~ sgn(X_2(\epsilon,\theta)) |X_2(\epsilon, \theta)| 
 \  = \  \bar {G}^{-1}_{ii} + K_{\sigma}\{\epsilon, <q_\nu> \} 
\end{equation}
with $K_{\sigma}$ being the coherent potential which has to be estimated self-consistently.

\begin{equation}
\bar{\lambda}_{\nu} = \lambda_{c \nu} + \lambda_{o \nu} + \lambda_{r \nu}
\end{equation}
in case of anodic current.

\begin{equation}
E^r_r = \sum_{\nu}
 \frac{\displaystyle \lambda_{r\nu}^2}{ \displaystyle \omega_\nu}; ~~
E^r_a = \sum_{\nu}
 \frac{\displaystyle \lambda_{a\nu}^2}{ \displaystyle \omega_\nu}; ~~
E^r_{ar} = 2 \sum_{\nu}
 \frac{\displaystyle \lambda_{r\nu} \lambda_{a\nu}}{ \displaystyle \omega_\nu}; ~~
 \end{equation}
are the reorganization energy for the reactant, adsorbate, and the cross reorganization energy, 
respectively.

\begin{equation}
\rho_r(\epsilon)  \, = \,   \sqrt{\frac{\beta }{4 \pi  E_r}} \, 
exp \left [ - \beta \frac {(\epsilon - \epsilon_r^{\prime} )^2 }{ 4 E_r} \right ]                
\end{equation}

Alternatively, we can also write

\begin{equation}
\epsilon_r^{\prime} = \epsilon^{0} _r  - \sum_{\nu} \frac{\lambda_{r \nu} \bar{\lambda}_{\nu}}{\omega_{\nu}} 
=    F^r_R - F^r_O - E_r^r  \equiv \eta - E^r_r 
\end{equation}

where
\begin{equation}
F^r_R = \epsilon_R - \sum^4_{\nu = 1}
 \frac{\displaystyle \lambda_{R\nu}^2}{ \displaystyle \omega_\nu}  -
 2\sum_{\nu = 1}^{4} \frac{\displaystyle \lambda_{R\nu} \lambda_{c\nu}} {\displaystyle \omega_\nu}
\end{equation}

\begin{equation}
F^r_O = \epsilon_O - \sum^4_{\nu = 1}
 \frac{\displaystyle \lambda_{o\nu}^2}{ \displaystyle \omega_\nu}  -
 2\sum_{\nu = 1}^{4} \frac{\displaystyle \lambda_{o\nu} \lambda_{c\nu}} {\displaystyle \omega_\nu}
\end{equation}

{\small $F_O$ and $F_R$ denote the free energies of the redox-couple in the oxidized and reduced states.
$\epsilon_R - \epsilon_O = \epsilon^o_r$, 
$\lambda_{R\nu} = \lambda_{r\nu} - \lambda_{o\nu}$. Thus $F_{R} - F_{O}$ gives the overpotential $\eta$ 
for the electron transfer reaction. Similarly, the fraction of overpotential drop between the electrode and 
adsorbate  is related to the change in the adsorbate free energy during the reaction
\begin{equation}
\epsilon_{a\sigma}^{\prime} = \epsilon^{0} _{a\sigma}  - \sum_{\nu} \frac{\lambda_{a \nu} \bar{\lambda}_{\nu}}{\omega_{\nu}} 
=    F^a_R - F^a_O - E_a^r  \equiv \alpha \eta - E^r_a + E^r_{ar}
\end{equation}

 }

\vspace{0.2in}
\hspace{0.2in}

{\small Rewriting the Anodic current expression in terms of overpotential, the expression for  $Q$ and $\rho_{r}$ takes the form as shown below. }

\begin{equation}
  Q^{an} \ = \  X_1(\epsilon,\theta)- \alpha \eta - E^{r} _{a} + E^{r} _{ar}   - \frac{\displaystyle (\epsilon - \eta +E^{r} _{r}) E^r_{ar} }
{\displaystyle 2 E^r_r} 
\end{equation}

\begin{equation}
  \rho^{an} _{r} (\epsilon) = \, = \,   \sqrt{\frac{\beta }{4 \pi  E^{r} _r}} \, 
exp \left [ - \beta \frac {(\epsilon - \eta + E^{r} _{r} )^2 }{ 4 E^{r} _r} \right ] 
\end{equation}

\vspace{0.2in}
\hspace{0.2in}

{\small Proceeding along similar lines of argument for the cathodic current, and noting that $\bar{\lambda}_{\nu} = \lambda_{c \nu} + \lambda_{o \nu} + \lambda_{a \nu}$ for cathodic current, the expression for $Q$ and $\rho_{r}$ obtained as shown below,}

\begin{equation} 
\mathrm{I}_{c} \, = \, 2\, e \theta |v_{ar}|^2 \sqrt{  \pi } \hbar^{-1}\, \int^{\infty}_{- \infty} 
sgn(X_2(\epsilon, \theta)) \, f(\epsilon) \, \, \rho_a (\epsilon) \, \rho_r(\epsilon)
d\epsilon 
\label{cathodeI}
\end{equation}

\begin{equation}
 Q^{cat} \ = \  X_1(\epsilon,\theta)- \alpha \eta + E^{r} _{a}    - \frac{\displaystyle (\epsilon - \eta - E^{r} _{r} +E^{a} _{ar}) E^r_{ar} }
{\displaystyle 2 E^r_r} 
\end{equation}

\begin{equation}
 \rho^{cat} _{r} (\epsilon) = \, = \,   \sqrt{\frac{\beta }{4 \pi  E^{r} _r}} \, 
exp \left [ - \beta \frac {(\epsilon - \eta - E^{r} _{r} +E^{r} _{ar} )^2 }{ 4 E^{r} _r} \right ] 
\end{equation}

\vspace{0.2in}
\hspace{0.2in}

{\small The coupling constants between adsorbate and various oscillator
modes are scaled by a factor $\sqrt{( 1- \theta^2)}$ to take into account the disolvation effect 
as  adlayer itself exhibits metallic properties 
in the higher coverage regime. Consequently, the  solvation and reorganization energy for the 
adsorbate  get scaled by a factor $( 1 - \theta^2)$, and the solvent induced cross  energy terms
are scaled as $\sqrt{( 1 - \theta^2)}$ \cite{}. No such scaling is present for solvation and reorganization
energies of the redox-couple. Thus the scaling laws for the various re-organisation are as follows }

\begin{equation} 
E^{r} _{ar} (\theta) = \sqrt{(1 - \theta^{2})} E^{r} _{ar} (0) \quad; \quad E^{r} _{a} (\theta) = (1 - \theta^{2}) E^{r} _{a} (0) 
\label{reorg1}
\end{equation}


\section{Numerical Results and Discussions}

\vspace{0.2in}
\hspace{0.2in}

{\small The basic concern in the article is toward current-overpotential characteristics with specific emphasis on the variation with the coverage factor 
($\theta$) and the fraction of overpotential drop ($\alpha \eta$) across the adsorbate. 
A first look at the expression for anodic current for a shows that the current is an overlap integral 
of three terms corresponding to the availability of vacant energy level at the electrode 
($1 - f(\epsilon)$), the density of states of the solvated redox couple $\rho^{\bf {an}}_{r}$ and the 
density of states of the adsorbate $\rho_{\bf {a}}^{\bf {an}}$.  
 The redox density of states has a Gaussian form in terms of $\epsilon$. 
The self-consistent evaluation of the coherent potential $k_{\sigma} (\theta)$ enforces a numerical derivation of the 
adsorbate density of states. However in the following limiting cases,  $k_{\sigma} (\theta) $ 
takes the value  

\begin{equation}
\lim_{\theta \rightarrow 0} k_{\sigma} = \epsilon -\frac {\displaystyle \epsilon - \epsilon_{a \sigma} - w_{ii}}
 {\displaystyle \theta} - w_{ii}
\end {equation}
and
\begin{equation}
\lim_{\theta \rightarrow 1} k_{\sigma} =  \epsilon_{a \sigma} 
\end {equation}
where
\begin{equation}
w_{ii} = \sum_k \frac {\displaystyle |v_{ik}|^2}{\displaystyle \epsilon - \epsilon_k}
\end{equation}.

Consequently, the adsorbate density of states can be analytically obtained in 
the limits $\theta \rightarrow$  0 and 1. 
Additionally, $ \epsilon^{\prime} _{a \sigma}  
 $ involved in performing the self-consistent evaluation of the coherent potential  takes the value 
{\bf as   $\alpha \eta - E_{a}(\theta)   +E_{ar}(\theta)$ for anodic current evaluation and 
$\alpha \eta + E_{a}(\theta) $ for} cathodic current estimation.  

\vspace{0.2in}

In what follows, we describe the current vs overpotential profile for different sets of parameters.
The adsorbate-electrode interaction is treated both in the weak ($v = 0.5 eV$) and strong  ($v = 2.0 eV)$
coupling limits. When the coverage is low, the adsorbate density of states has a single peak Fig. \ref{andos} . An important consequence of the strong coupling limit is the splitting of the adsorbate level 
in bonding and anti-bonding states for low $\theta$ Fig.\ref{strandos}. 
This feature is recaptured in the present analysis since
energy dependence of $\Delta(\epsilon)$ is explicitly treated in the present approach (Eqs ~\ref{enerperp}, ~\ref{enerpar}).
On the other hand, the well known wide-band approximation for $\Delta(\epsilon)$  fails to 
provide the bonding anti-bonding splitting. In the monolayer regime, due to the
2-d bond formation by the adsorbate layer, its density of states acquires a flat profile, irrespective of the strength of the electrode-adsorbate coupling (Fig ~\ref{andos}, ~\ref{strandos}) . 
The table I summarizes the values of parameters used in the calculations.

\begin{table}[ht]
 \caption{Values of parameters used in calculation in eV}
\centering
\begin{tabular}{c c c c c c c c}
\hline \hline
   &  $v$ & $\Delta_{||}$ & $\Delta_{\perp}$ & $\mu$ & $E_{r}$ & $E_{ar}$ (0) & $E_{a}$ (0) \\
\hline
strong & 2.0 & 0.75 & 1.5 & 4.5 & 1.0 & 0.25 & 0.75 \\
weak   & 0.5 & 0.75 & 1.5 & 4.5 & 0.6 & 0.2 & 0.4 \\
\hline
\end{tabular}
\label{table:param}
\end{table}

\vspace{0.2in}
\hspace{0.2in}

{\small Ideally, under zero overpotential condition, the anodic and cathodic currents  are
 expected   to be equal in magnitude.
 This implies that the profile of the product $ \rho_{r} ^{an} (\epsilon) * \rho_{a} ^{an} (\epsilon) $ 
for anodic current  is identical to the product profile 
$ \rho_{r}^{cat} (\epsilon) * \rho_{a} ^{cat} (\epsilon)$  for 
the cathodic current.  This is a consequence of 
equal separation between the peak positions of adsorbate and reactant  density of states for anodic and 
cathodic processes during equlibrium.   [Fig. \ref{anoverlap},~\ref{catoverlap}]. The corresponding plots for strongly coupled regime is also shown in Fig. \ref{sanoverlap},~\ref{scatoverlap} 

\vspace{0.2in}

As noted earlier, the  electrochemical   
potential $\mu$  has been set as the zero of energy scale for the direct 
electron transfer reaction. The presence of additional charge particles for the bridge assisted
electron transfer reaction, namely the adsorbates, changes the equilibrium  potential of the electrode.
This is turn gets reflected as a $\theta$ dependent  variation  $\Delta \phi (\theta) $ in $\mu (\equiv 0)$.}
The fact that the anodic and cathodic currents at equilibrium potential are identical in magnitude 
provides a  novel method for the  determination of  $\Delta \phi(\theta)$. Thus 
the relation $I_a(\eta =0) = I_c(\eta = 0)$
with $f(\epsilon)   = {\bf  (1 + exp(-\beta (\epsilon + \Delta \phi(\theta)))^{-1}}$
\{ \} (cf eqs. ~\ref{anodeI} and ~\ref{cathodeI}) enables us to evaluate $\Delta \phi(\theta)$.
The  variation of 
$\Delta \phi$ with respect to $\theta $ is shown in Fig.\ref{thetas} in the limit of weak and strong 
adsorbate-electrode  interaction, with $E^{r} _{r}$ = 0.6 eV, $E^{r} _{ar}$(0) = 0.2 eV, $E^{r} _{a}$ (0) = 0.4 eV.
The value of  $\Delta \phi(\theta)$ depends on the strength of
coupling $v$; its magnitude  increases as the coupling becomes stronger. $|\Delta \phi(\theta)|$
is again large for low $\theta$ values and remains almost constant in this region. Note that in this regime, the  
charge on the adsorbate remains localized on the adsorption site. $|\Delta \phi(\theta)|$ starts
diminishing sharply for $\theta  > 0.6$ and it tends to 0 as $\theta \rightarrow 1$. This behaviour
is expected. As $\theta \rightarrow 1$, the adsorbate layer becomes metallic and gets incorporated 
in the electrode. The electron transfer acquires the characteristics of a direct heterogeneous
reaction, and consequently as noted earlier, the electrochemical potential $\mu$ again lies at the
zero of the energy scale.

\vspace{0.2in}
\hspace{0.2in}

{\small We first present the current-overpotential profile in the
weak coupling limit ($v = 0.5 ~ eV$) for a range of $\theta$ and $\alpha$.  The employed values of various 
reorganization energies are  $E^{r} _{r}$ = 0.6, $E^{r} _{ar}$(0) = 0.2, $E^{r} _{a}$ (0) = 0.4. 
The general behaviour can be analysed by looking at the  case of lower coverage and high 
coverage regimes respectively, and then by investigating the effect of variation of $\alpha$ 
in these limits. 
Fig. \ref{Ivseta}  shows that for a fixed $\alpha$, anodic current as well as 
the current peak height increases with $\theta$ in  the small $\theta$ range (curve a and b).
This feature arises due to a better overlap between the reactant and adsorbate density of states,
whose peak positions are approximately separated by a distance  $E_r^r + E_a^r(\theta) - 
E_{ar}^r(\theta)$. An increase in $\theta$ reduces $E_a^r$ and $E_{ar}^r$ (cf eq. ~\ref{reorg1}), and hence
the peak separation diminishes and the overlap gets enhanced. The presence of anodic current 
peak at $\eta_p$ signifies negative differential resistance for $\eta > \eta_p$. This feature 
is absent in the higher coverage limit. For large value of $\theta$, the current at 
higher $\eta$ exhibits a saturation effect. This is a consequence of the fact that the maximum
n
in the adsorbate density of states $\rho^{an}_a $ is now absent. $\rho^{an}_a $ now acquires a plateau
profile (Fig \ref{andos}). The plateau height, and therefore the overlap between the reactant and adsorbate 
density of states decreases  with the increasing coverage . Therefore a decrease 
in the saturation current results as $\theta \rightarrow 1$ (curve $\theta $ = 0.7 and 0.9 in Fig. \ref{Ivseta}). }

\vspace{0.2in}

The effect of the $\alpha$ variation on the anodic current is highlighted in Fig. \ref{Ivt1},~\ref{Ivt3} and ~\ref{Ivt7}
The effect is more pronounced in the low coverage regime due to the presence of adsorbate
density of states peak. The reactant and adsorbate density of states peak separation increases
with the increasing $\alpha$. Consequently, the maximum overlap between the two occurs at
larger $\eta$. This explains the occurrence of the  anodic current peak  at higher $\eta$
values as $\alpha$ increases . On the other hand, the near constant
adsorbate density of states for large $\theta$ ensures a minimal effect of $\alpha$ variation
on the anodic current (Fig. \ref{Ivt7},~\ref{Ivt9}). 

\vspace{0.2in}

 {\small   Next the strong coupling limit with  
\lbrack $v$ = 2.0 eV, $E^{r} _{r}$ = 1.0, 
$E^{r} _{ar}$(0) = 0.25, $E^{r} _{a}$ (0) = 0.75 \rbrack is considered.
Figures  \ref{Iact1},\ref{Iact3},\ref{Iact7} and \ref{Iact9} shows the current overpotential response in the 
strong coupling regime. As in the case of low coverage, the $I_a $ {\it vrs} $\eta$ plot
exhibits a negative-differentail region (Fig. \ref{Iact1},~\ref{Iact3}). } 

\vspace{0.2in}

{\small More importantly, the presence of two peaks in $\rho_a^{an}$ when coupling $v$ is large and $\theta$ is small
(Fig. \ref{strandos}) leads to a saddle point and a maximum in the $I_a$ vrs. $\eta$ plot. For the set of parameters currently employed,
the $I_{A,Max}$ now occurs at a much larger $\eta$ in comparison to the weak coupling limit,
and may not be accessible experimentally. 
However, the saddle point in the current appears in an overpotential range where
the anodic current peak appears in the weak coupling limit. For large coverage, current
potential profile are  similar in strong and weak coupling limit. Interestingly,
the saturation current is smaller in the large coupling case due to a decrease
in the height of  $\rho^{an}_a $. In fact this lowering of the current
in the strong coupling is holds true for any coverage and $\eta$. This is
shown in Fig. \ref{ci0} wherein the variation of equilibrium current $I^o$ with
respect to coverage is plotted  The $I^o$ is smaller for larger $v$, and as explained
earlier in the context of Fig. \ref{Ivseta}, shows a maximum in the intermediate coverage regime.
However it may be noted that when $v \rightarrow 0$, current would be proportional to
$|v|^2$, and an increase in $v$ in this very weak coupling limit will lead to an
increase in the current. 

\vspace{0.2in}

The results presented till now correspond to anodic current. But the formalism developed here
also  yields the  cathodic current. In fact the equivalence of anodic and cathodic currents 
at the equilibrium potential has been earlier employed to determine the the variation
in the equilibrium potential due to varying adsorbate coverage. The dependence of the cathodic
current $I_{c}$ on overpotential $\eta$ is plotted in Fig. (\ref{Iancat}). It is often presumed that
$I_c = e^{-\eta} I_a$ (Fig. \ref{Iancat}). The present `microscopic' calculations show that
it is not entirely true. The calculated current value is slightly larger than the $e^{-\eta} I_a$
when $\eta < $ 0. 
  
\vspace{0.2in}

{\small The high coverage regime of $\theta \rightarrow 1$ corresponding to a formation of monolayer of 
A decrease in the current for higher $\eta$ when the coverage is low virtually mimics
the Marcus inverted region for a homogeneous electron transfer reaction. On the other hand,
the current getting saturated at higher $\eta$ when the coverage is large is also true
for a  direct heterogeneous electron transfer reaction.  Thus depending on
the extent of coverage, an adsorbate mediated electron transfer at an electrode 
exhibits  the characteristics of both homogeneous and heterogeneous
electron transfer reactions. The localization of adsorbate electron  at low
coverage and its delocalization at high coverage is the reason behind this
phenomena. } }

\section{Summary and Conclusions}

\vspace{0.2in}
\hspace{0.2in}

{\small In this work, we considered electron transfer in an electrochemical system, from a solvated redox to an electrode 
mediate by intervening adsorbate atoms. Further randomness is introduced in the model in terms of the 
coverage factor which relates to the number of adsorbate atoms adsorbed on the electrode surface. 
The theory developed is valid for a range of regime, lone adsorbate mediate transfer to the monolayer formated direct 
electron transfer regime. The inherent randomness involved in the adsorbate distribution 
on the surface has been tackled by coherent potential approximation (CPA) and separate expression 
are derived for anodic and cathodic current.}

\vspace{0.2in}
\hspace{0.2in}

{\small Explicit attention was paid to the low coverage and high coverage regime, even though 
the formalism is valid for all regime, since at these two regions the theory could be compared with 
pre-existing literature. Plots were also provided for intermediate regimes and additionally, the 
effect of the adsorbed atoms on the Fermi level of the electrode 
were incorporated by means of a shifted potential $\Delta \phi (\theta)$, ensuring that the anodic and cathodic current 
were equal under zero overpotential condition. 

\vspace{0.2in}
\hspace{0.2in}

{\small  The analysis also provides a novel method for determining the
variation in  $\Delta \phi (\theta)$ with changing adsorbate coverage.}

\vspace{0.2in}
\hspace{0.2in}

{\small  The fraction of overpotential drop across the electrode-adsorbate is incorporated and the collective 
plots are analysed. We have proved that this fraction of overpotential drop plays a significant role 
in determining the response behaviour of current, typically the location and extent of the maximas 
in case of lower coverage situations. while in case of high coverage regime, the effect is 
not profound and the electron transfer follows the traditional direct electron transfer as expected 
from heuristic arguments. }

\vspace{0.2in}
\hspace{0.2in}

{\small  The dependence of anodic current  in the weak and strong electrode-adsorbate coupling 
is analyzed. In the former case, $I_a$  vrs overpotential profile exhibits a peak, where as in the 
later case, and in the same overpotential region, the current plot shows a saddle point behaviour.
This fact can be used to distinguish a weakly chemisorbed bridge from a strongly chemisorbed one.
These distinguishing features occur only when the coverage is low. At high coverage, $I_a \sim  \eta$
plots have identical profile for weak and strong coupling cases}

\vspace{0.2in}
\hspace{0.2in}

{\small The calculated cathodic current gives a slightly higher value of $I_c$ in comparison
to a presumed $I_c$ which equals  $e^{-\eta} I_a$.}

\vspace{0.2in}
\hspace{0.2in}

{\small  At low coverage, it is possible to recover the Marcus inverted region, which is absent 
when the coverage is large. The localized  nature of the adsorbate orbital when coverage is low, and 
its getting delocalised for high coverages leads to this behaviour. }

\newpage
\section*{appendix}
\renewcommand{\theequation}{A-\arabic{equation}}
\setcounter{equation}{0}

\vspace{0.2in}
\hspace{0.2in}

The microscopic current associated with the electron transfer reaction depends on the average
value of the rate of change of electronic occupancy of the redox orbital [   ]

\begin{equation}
I = -e \left < \frac{\displaystyle \partial  n_r}{\displaystyle \partial  r} \right >
\end{equation}

Treating $v_{ar}$ as a small quantity, a linear  response formalism can be used to
evaluate $< \dot {n}_r>$. Consequently,

\begin{equation}
I \ = \ \frac{e}{\hbar^2} \sum_{\sigma} \int^{\infty}_{-\infty} 
 < [ V^\dagger_{I\sigma }(0), V_{I\sigma}(t) ]_- > dt
\end{equation}
where

\begin{equation}
V_{I\sigma } \ = \ v_{ar} C_{a\sigma}^\dagger C_{r} 
\end{equation}
The first term in the commutator leads to anodic and the second one
gives the cathodic current.
The expectation value in (A-2) now corresponds to a density matrix
defined for the Hamiltonian $H^{\prime} = H - \sum_{\sigma} (V^\dagger_{Ir} +
V_{Ir})$.  $H^\prime$ also determines the time evolution of various operators in (A-2).
Employing the Frank-Condon approximation and treating the low frequency 
polarization modes in the semi-classical approximation, anodic current 
is obtained as 
\begin{equation}
I_a \ = \ \frac{e}{\hbar}^2  \int^\infty_{-\infty} dt
|V_{at}|^2 << C^\dagger_{r}(0) C_{r}(t) >_F
< C_{a\sigma} (0) C^\dagger_{a\sigma} (t) >_F >_B 
\end{equation}
where
\begin{equation}
< c^{\dagger}_{r} (0) c_{r} (\tau) >_F   
\ = \  \frac{\displaystyle 1}{\displaystyle \pi}\int^\infty_{-\infty} 
 e^{-i\epsilon\tau/
{\hbar}} \delta(\epsilon - \epsilon_r ) d\epsilon 
\end{equation}
The time correlation function involving $c_a, ~ c^{\dagger}_a$ can be expressed
in terms of adsorbate Green's function
\begin{equation}
< c_{a\sigma} (0) c^\dagger_{a\sigma} (t) >_F   
\ = \  \frac{\displaystyle 1}{\displaystyle \pi}\int^\infty_{-\infty}(1 - f(\epsilon)) e^{i\epsilon\tau/
{\hbar}} (Im G_{ii})_{i=a} d\epsilon 
\end{equation}
where
\begin{equation}
(G_{ii}(\epsilon))_{i = a} \ = \ < 0| c_{a\sigma}<\frac{\displaystyle 1}{\displaystyle \epsilon - H^{\prime}}>_{c,i = a}
c^{\dagger}_{a\sigma}|0>_F  
\end{equation}
Here $<...>_F$ implies an average over electronic degrees of freedom,
keeping bosonic variables as fixed parameters.  $<...>_B$ denotes the
thermal average over boson modes.  
$<...>_{c,i = a}$ denotes  a restricted configuration average.  It implies
that while obtaining the configuration average, the site a, which is
occupied by an adsorbate and through which the electron transfer takes place, 
is excluded from the averaging.  The occupancy
status of the remaining sites are still unspecified.  We replace the
random medium encompassing the remaining sites by an effective medium 
using the CPA technique.  The picture which now emerges is the one in
which a reactant is coupled to an adsorbate occupying the site a, and this 
particular adsorbate is embedded in a two dimensional effective medium.

\vspace{0.2in}
\hspace{0.2in}

The randomness associated with these  sites  can be handled using the 
coherent potential approximation [   ]. 
Accordingly, the  random energy operator 
$\hat \epsilon_{i\sigma}(<q_\nu>) n_{n\sigma}$  in eq. ~\ref{esite} is now replaced by a deterministic operator
$k_{\sigma} n_{i\sigma}$.  The coherent potential $k_{\sigma}(\epsilon, <q_{\nu}> )$
is same for all the sites, but
depends on the energy variable $\epsilon$.  $k_{\sigma} $ is
determined self-consistently through the expression [ ]

\begin{equation}
\bar{G}_{ii} \ = \ \frac{1}{N_{||}} \sum_u \frac{1}{\epsilon-k_{\sigma}(\epsilon, <q_\nu>) - \epsilon_u - 
W'(\epsilon,u)} \ = \ \frac{1-\theta}{\epsilon_{a\sigma} - k_{\sigma}(\epsilon, <q_\nu> )}  
\end{equation}
 where 
 2D adsorbate lattice has $N_{||}$ number  of sites, and 

\begin{equation}
 \begin{array}{lcl}
W(\epsilon,u) & = & \sum_j e^{i{\bf u}. {\bf R}_{ji}} [v_{ij} + W'_{ij} (\epsilon)] \\
\\
& = & \epsilon_u + \sum_j e^{i{\bf u}. {\bf R}_{ji}} W'_{ij} (\epsilon) \\
\\
& = & \epsilon_u + W'(\epsilon,u) \end{array} 
\end{equation}

The   $(G_{ii})_{i=a}$   can  be related to the complete configuration
averaged GF $\bar{ G_{ii}}$ 

\begin{equation}
(G_{ii})_{i=a} \ = \ \bar {G_{ii}}(\epsilon) [1 - (\hat{\epsilon}_{a\sigma}(q_\nu ) - k_{\sigma}(\epsilon))
\bar G_{ii}(\epsilon)]^{-1} 
\end{equation}

	The tedious summation over the momentum k of metal states and
the momentum u of the Bloch states in 2D adsorbate layer, commensurate
with the underlying electrode surface lattice, can be considerably
simplified  under the following assumptions. (i)
The separability of the metal state energy $\epsilon_k$    in the
direction parallel and perpendicular to the surface. (ii) The substrate
density of states in the direction perpendicular to surface is taken to
be Lorentzian, whereas the same is assumed to be rectangular along the
surface. (iii) The adsorbate occupies the `on-top' position on the electrode, and
is predominantly coupled to the underlying substrate atom with coupling strength $v$.
Consequently, eq.(A9) now becomes

\begin{equation}
\begin{array}{lcl}
\bar{G}_{ii} (\epsilon) & = & \frac{ \displaystyle 1-\theta}{ \displaystyle \epsilon_{a\sigma}\{<q_\nu>\}
 - k_\sigma \{\epsilon, <q_\nu>\}} \\ 
\\
& =  & \frac{ \displaystyle 1}{\displaystyle 2\Delta_{||}  (B - A) \mu} \left[ (A-C) \ln \left(
\frac{A-\Delta_{||}}{A+\Delta_{||}}\right) - (B-C) \ln \left(
\frac{B-\Delta_{||}}{B + \Delta_{||}}\right) \right] 
\end{array} 
\label{selfconeq}
\end{equation}
 with 

\begin{equation}
A/B \ = \ \frac{1}{2} [(C+D) \pm \{(C+D)^2 - 4(CD - \frac {v^2}{\mu}) \}^{\frac{1}{2}}] 
\end{equation}

\begin{equation}
C \ = \ \frac{\epsilon - k_\sigma\{\epsilon, <q_\nu>\}}{\mu} \quad; \quad \ D \ = \ \epsilon - i \Delta_{\perp};
\quad \mu \ = \ \theta \delta /\Delta_{||}
\end{equation}

$\delta$  is the half-bandwidth of adsorbate monolayer and $\mu$ is the same for an arbitrary coverage $\theta$.
$2 \Delta_{||} $ is the substrate bandwidth at the surface and $ \Delta_{t}$ the total bandwidth of the substrate 
$ 2 (\Delta_{||} + \Delta_{\perp} ) = \Delta_{t} $. 

\begin{equation}
\frac{1}{N_{\perp}} \sum \delta (\epsilon - \epsilon_{k_{z}}) = \frac{1}{\pi} \frac{\Delta_{\perp}}{(\epsilon - \epsilon_{k_{z}})^{2} + \Delta_{\perp} ^{2}}
\label{enerperp}
\end{equation}

\begin{equation}
\frac{1}{N_{||}} \sum \delta(\epsilon - \epsilon_{u}) = 1/2\Delta_{||} \quad; \quad -\epsilon < \Delta_{||} < \epsilon
\label{enerpar}
\end{equation}

Where the separability of metal states energies ${\epsilon_{k}}$ in directions parallel and perpendicular 
to the surface is a crucial assumption made here. $\epsilon_{k} = \epsilon_{u} + \epsilon_{k_{z}} $.

The self-consistent value of $k_{\sigma}(\epsilon)$  is obtained from eq.~\ref{selfconeq}, which also determines $\bar{G}_{ii}$.

Expressing
\begin{equation}
\bar {G}^{-1}_{ii} + K_{\sigma}\{\epsilon, <q_\nu> \} \ = \ X_1(\epsilon, \theta) + i~ sgn(X_2(\epsilon,\theta)) |X_2(\epsilon, \theta)|
\end{equation}
enables us to write (cf. eq.(A11))

\begin{equation}
Im(G_{ii})_{i=a} \ = \ -sgn(X_2)\frac{\displaystyle |X_2|}{\displaystyle (X_{1} (\epsilon,\theta) - \hat {\epsilon}_{a\sigma}
(q_\nu) )^2 + X^2_2}
\end{equation}
where  sgn$(X)$  = +1 when  $X > 0 $ and -1 otherwise.

	To evaluate the anodic current $I_a$, one finally needs to
carry out the thermal average over low frequency boson modes.
The required density matrix for this average in the
semi-classical limit is 

\begin{equation}
P(q_\nu ) \ = \ W(q_\nu )/\int^{\infty}_{-\infty} W(q_\nu ) dq_\nu    
\end{equation}

\begin{equation}
W(q_\nu ) \ = \ exp[-\beta\sum_{\nu=1,2} \frac{\omega_{\nu}}{2} (p_{\nu}^2 + q_{\nu}^2)
+ \bar{\lambda}_{\nu} q_{\nu}] 
\end{equation}

\vspace{0.2in}
\hspace{0.2in}

{\small With the above defined probability function the net expression for anodic current is shown below}

\begin{eqnarray} 
I_{a} & = &e \langle n_{r} \rangle \mid \upsilon _{ar} \mid ^{2} \frac{1}{\sqrt{\pi} \hbar} \frac{1}{\mathrm{Z}} \int d \epsilon
\int (\Pi _{\nu} dq_{\nu} )\int dt \int d\tau \lbrack (1 - f(\epsilon) \rbrack \nonumber \\
& & exp[-\beta\sum_{\nu=1,2} \frac{\omega_{\nu}}{2} (p_{\nu}^2 + q_{\nu}^2) + \bar{\lambda}_{\nu} q_{\nu}]  \quad  exp[i (\epsilon - \epsilon_{r} - \sum_{\nu} \lambda_{r \nu} q_{\nu}) t] \nonumber \\
&  &exp[i(X_{1} (\epsilon, \theta) - \epsilon_{a} - \sum_{\nu} \lambda_{a \nu} q_{\nu} ) \tau - \mid X_{2} (\epsilon, \theta) \mid \mid \tau \mid ] 
\end{eqnarray}

\vspace{0.2in}
\hspace{0.2in}

{\small where $\mathrm{Z} = \int (\Pi _{\nu} dq_{\nu} ) exp[ -\beta \sum_{\nu} \frac{\omega_{\nu}}{2} (p_{\nu}^2 + q_{\nu}^2) + \bar{\lambda}_{\nu} q_{\nu}] $  The expression for cathodic current has a similar form with the $ 1 - f(\epsilon) $ replaced by $f(\epsilon)$ and $\bar{\lambda}_{\nu}$ defined accordingly. For anodic current $\bar{\lambda} _{\nu} = \lambda_{c \nu} + \lambda_{o \nu} + \lambda_{r \nu} $, while for cathodic current, $\bar{\lambda} _{\nu} = \lambda_{c \nu} + \lambda_{o \nu} + \lambda_{a \nu}$. Carrying out the various integrations leads to the result employed in the main article.}

\begin{equation}
 I_{a} = e \langle n_{r} \rangle \mid \upsilon _{ar} \mid ^{2} \frac{1}{\sqrt{\pi} \hbar} \int d \epsilon (1 - f(\epsilon)) \rho_{a} (\epsilon) \frac{\mathrm{Re} \omega_{z}}{2 \sqrt{\pi \mathrm{P}}} 
\end{equation}


\newpage
\subsection*{Figures}

\begin{figure}[htb]
 \begin{center}
  \includegraphics[scale=0.6,angle=-90]{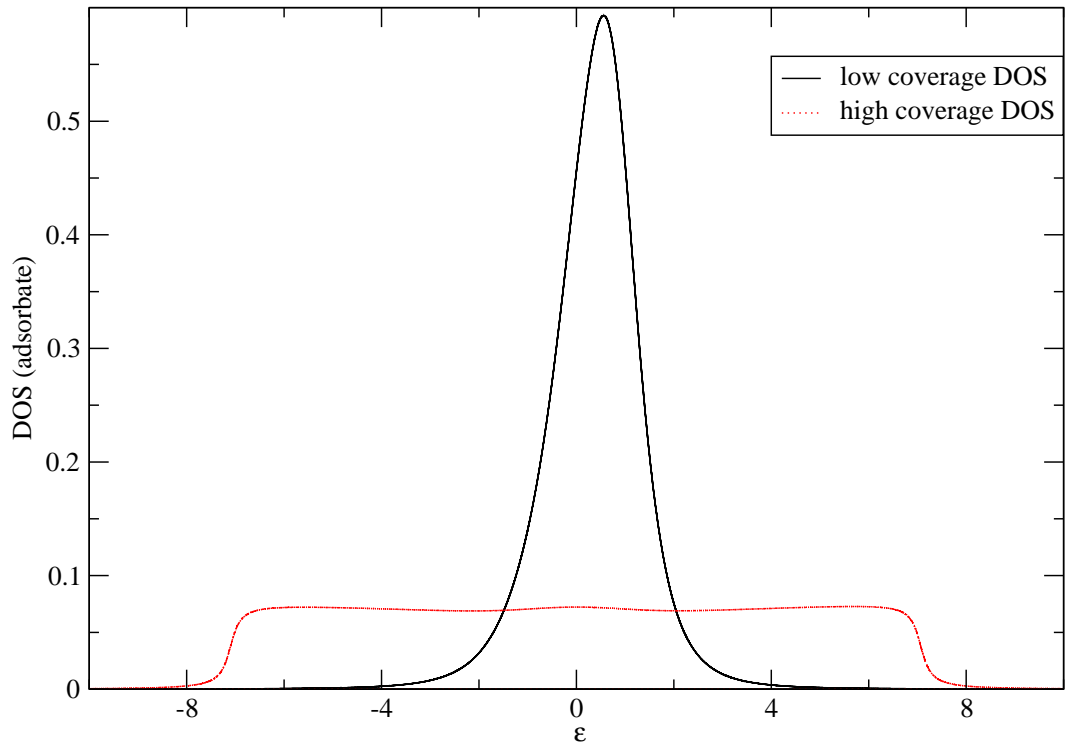}
\caption{Comparison of density of states of the adsorbate for weakly coupled regime at low ($\theta$ = 0.1) and high coverage factor ($\theta$ = 0.9). he values of parameters (in eV) are as follows: $E^{r} _{r} = 0.6, E^{r} _{ar}(0) = 0.2, E^{r} _{a} = 0.4$ and $v$ = 0.5 eV. }
\label{andos}
 \end{center}
\end{figure}

\begin{figure}[htb]
\begin{center}
 \includegraphics[scale=0.7,angle=-90]{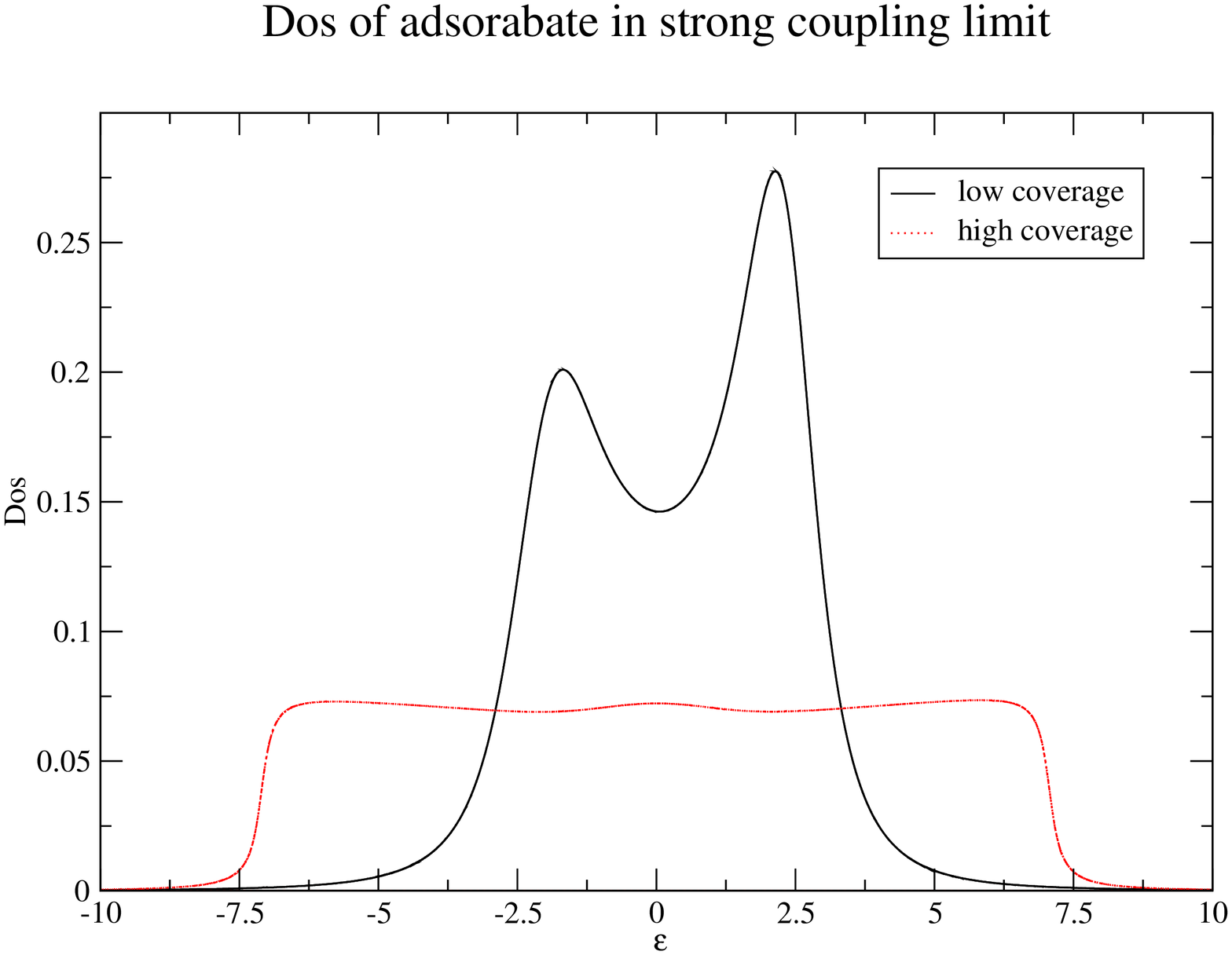}
\caption{Comparison of density of states of adsorbates for strong coupling regime at low and high coverage factor. The values of the various parameters employed (in eV) are as follows: $E^{r} _{r} = 1.0, E^{r} _{ar}(0) = 0.25, E^{r} _{a} (0) = 0.75, \Delta_{||} = 1.5, \Delta_{\perp} = 1.5, \mu = 4.5,\upsilon = 2.0$ }
\label{strandos}
\end{center}
\end{figure}

\begin{figure}[htb]
 \begin{center}
  \includegraphics[scale=0.6,angle=-90]{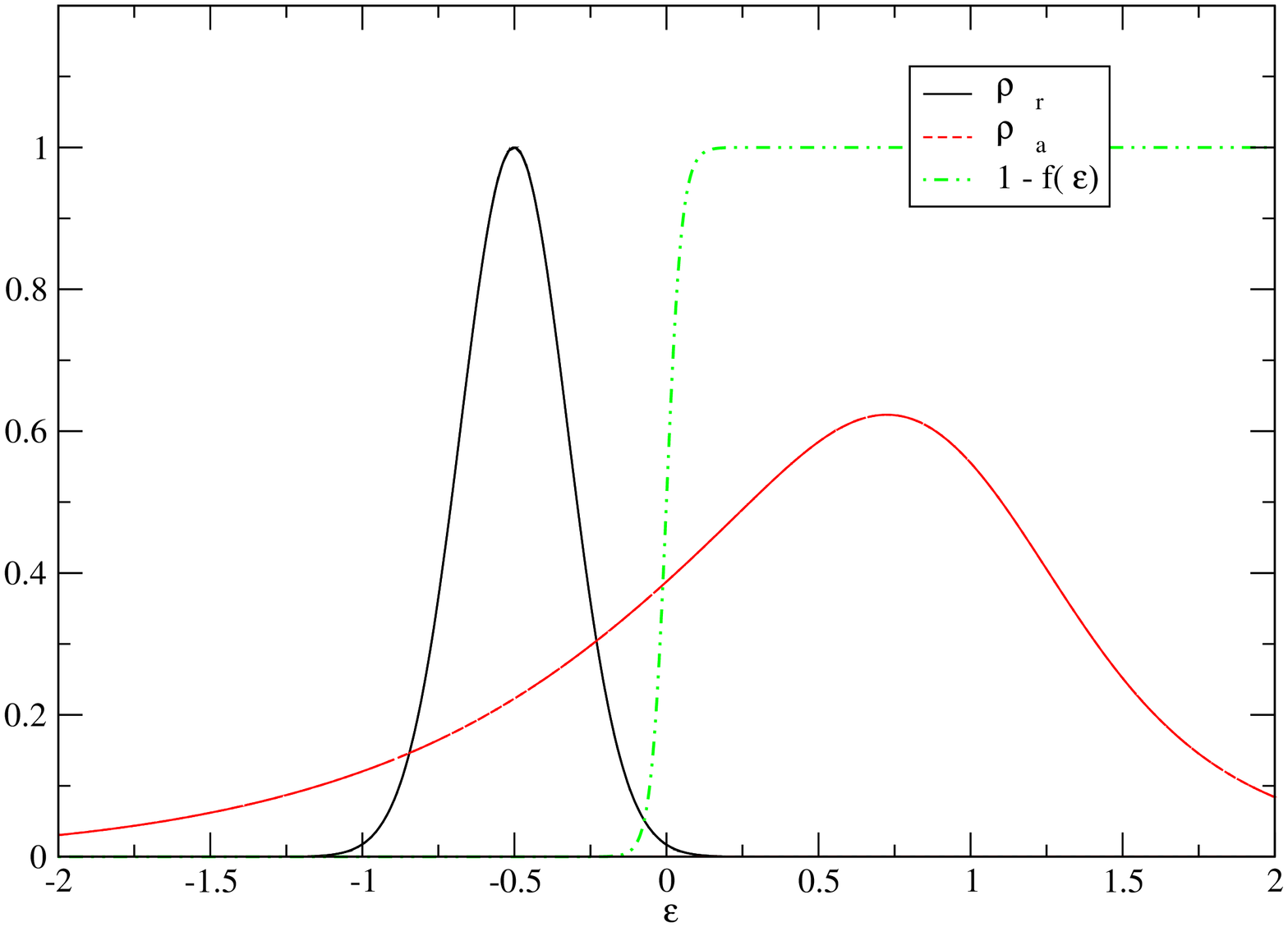}
\caption{Plots showing the  density of states for redox, adsorbate and the Fermi distribution for anodic current under zero overpotential. The weakly coupled regime and low coverage of $\theta$ = 0.3 is considered here .The values of parameters (in eV) are as follows: $E^{r} _{r} = 0.6, E^{r} _{ar}(0) = 0.2, E^{r} _{a} = 0.4$ and $v$ = 0.5 eV . }
\label{anoverlap} 
\end{center}
\end{figure}

\begin{figure}[htb]
 \begin{center}
  \includegraphics[scale=0.6,angle=-90]{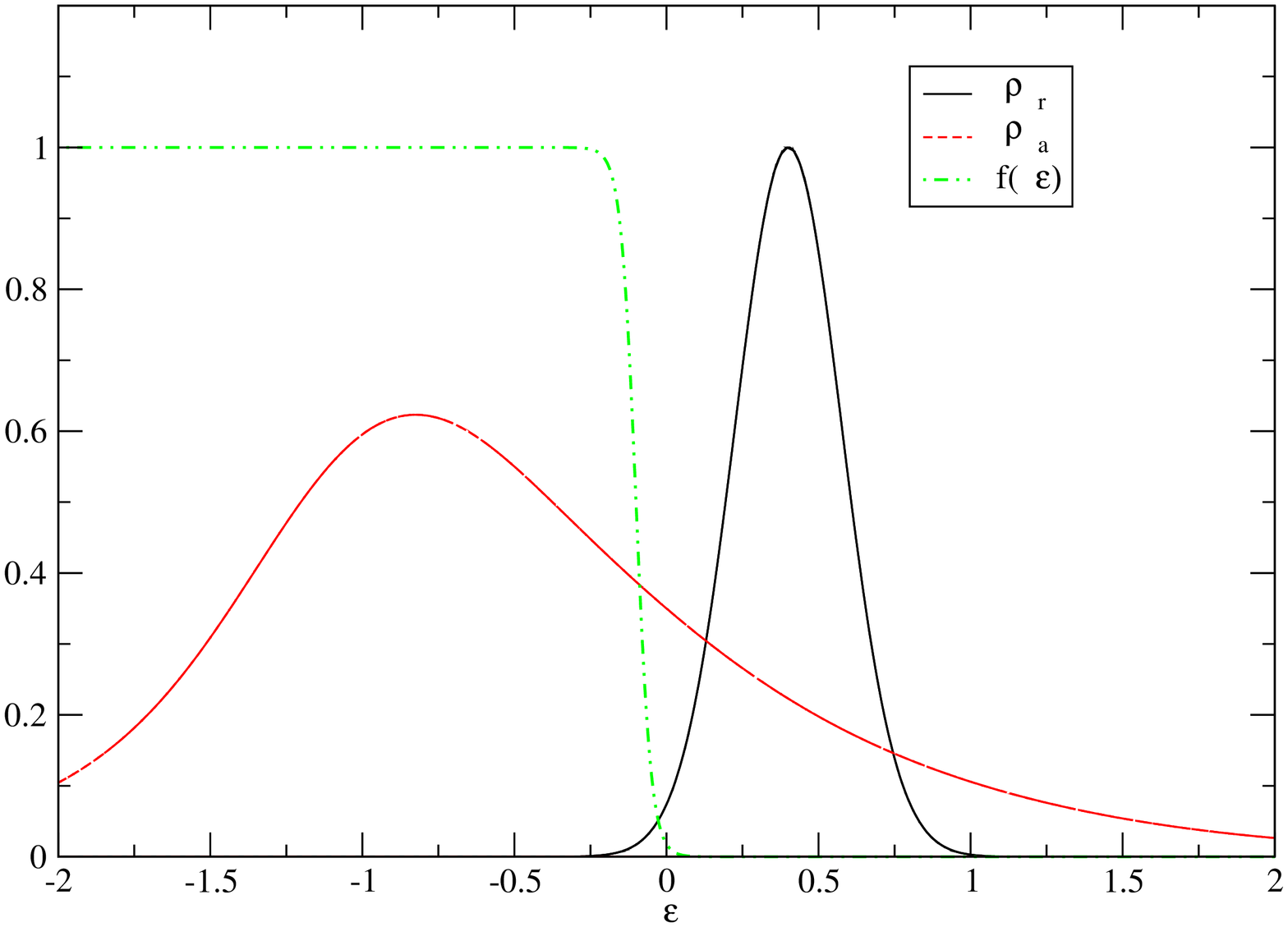}
\caption{Plots showing the density of states for redox, adsorbate and Fermi distribution for cathodic current at zero overpotential. The values of parameters are same as in \ref{anoverlap} }
\label{catoverlap} 
\end{center}
\end{figure}

\begin{figure}[htb]
\begin{center}
 \includegraphics[scale=0.7,angle=-90]{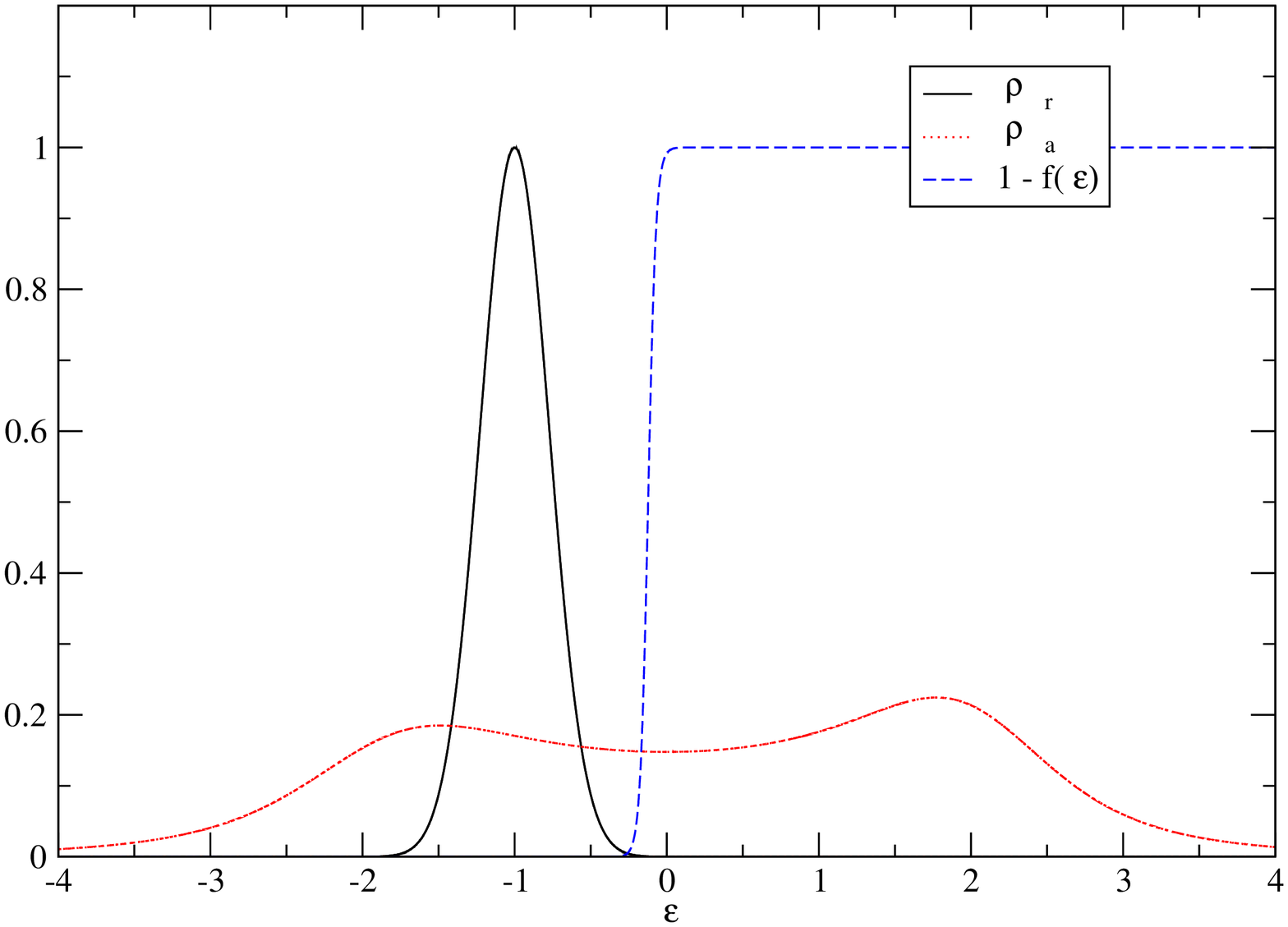}
\caption{Plots showing the  density of states for redox, adsorbate and the Fermi distribution for anodic current under zero overpotential. The strongly coupled regime and low coverage of $\theta$ = 0.3 is considered here .The values of parameters (in eV) are as follows: $E^{r} _{r} = 1.0, E^{r} _{ar}(0) = 0.25, E^{r} _{a} = 0.75$ and $v$ = 2.0 eV .}
\label{sanoverlap}
\end{center}
\end{figure}

\begin{figure}[htb]
\begin{center}
 \includegraphics[scale=0.7,angle=-90]{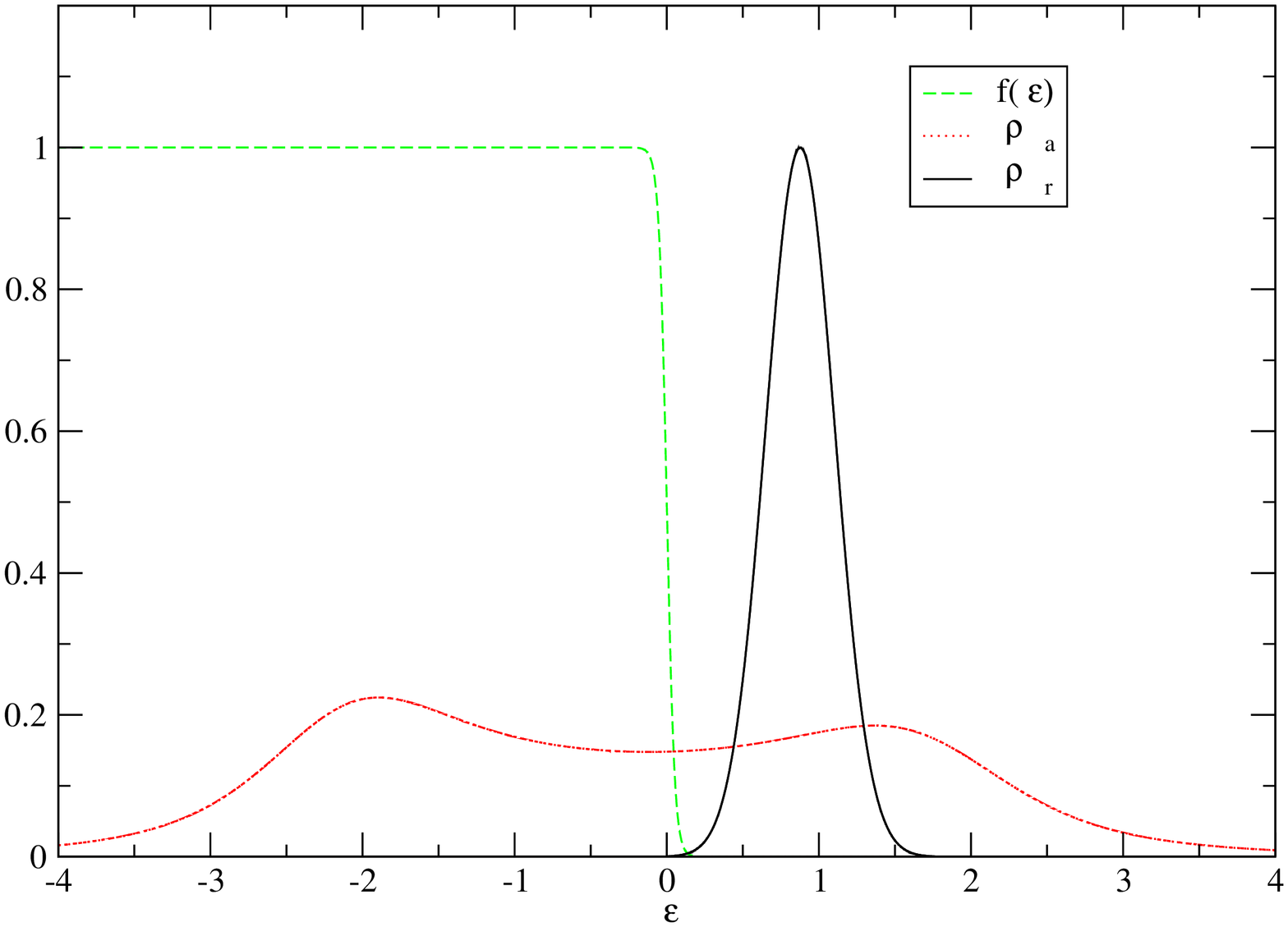}
 \caption{Plots showing the density of states for redox, adsorbate and Fermi distribution for cathodic current at zero overpotential. The values of parameters are same as in \ref{sanoverlap} }
\label{scatoverlap}
\end{center}
\end{figure}

\begin{figure}[htb]
\begin{center}
 \includegraphics[scale=0.7,angle=-90]{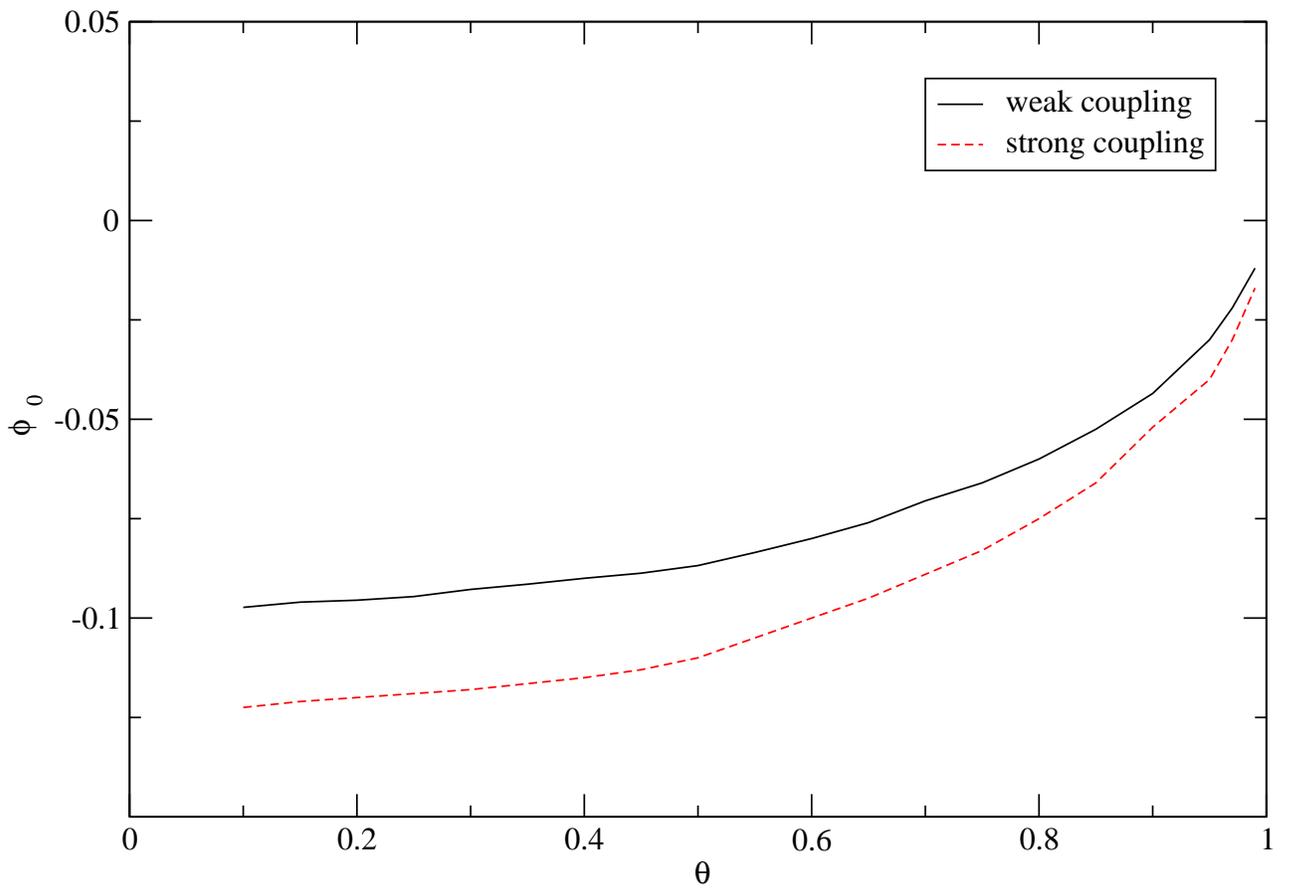}
\caption{Plots showing the variation of $\Delta \phi$ with respect to $\theta$ the coverage factor. The values of re-organisation energies employed were same in both the curves. $E_{r}$ = 0.6 eV, $E_{a} (0) $ = 0.4 eV, $E_{ar} (0) $ = 0.2 eV}
\label{thetas}
\end{center}
\end{figure}

\begin{figure}[htb]
 \begin{center}
  \includegraphics[scale=0.6,angle=-90]{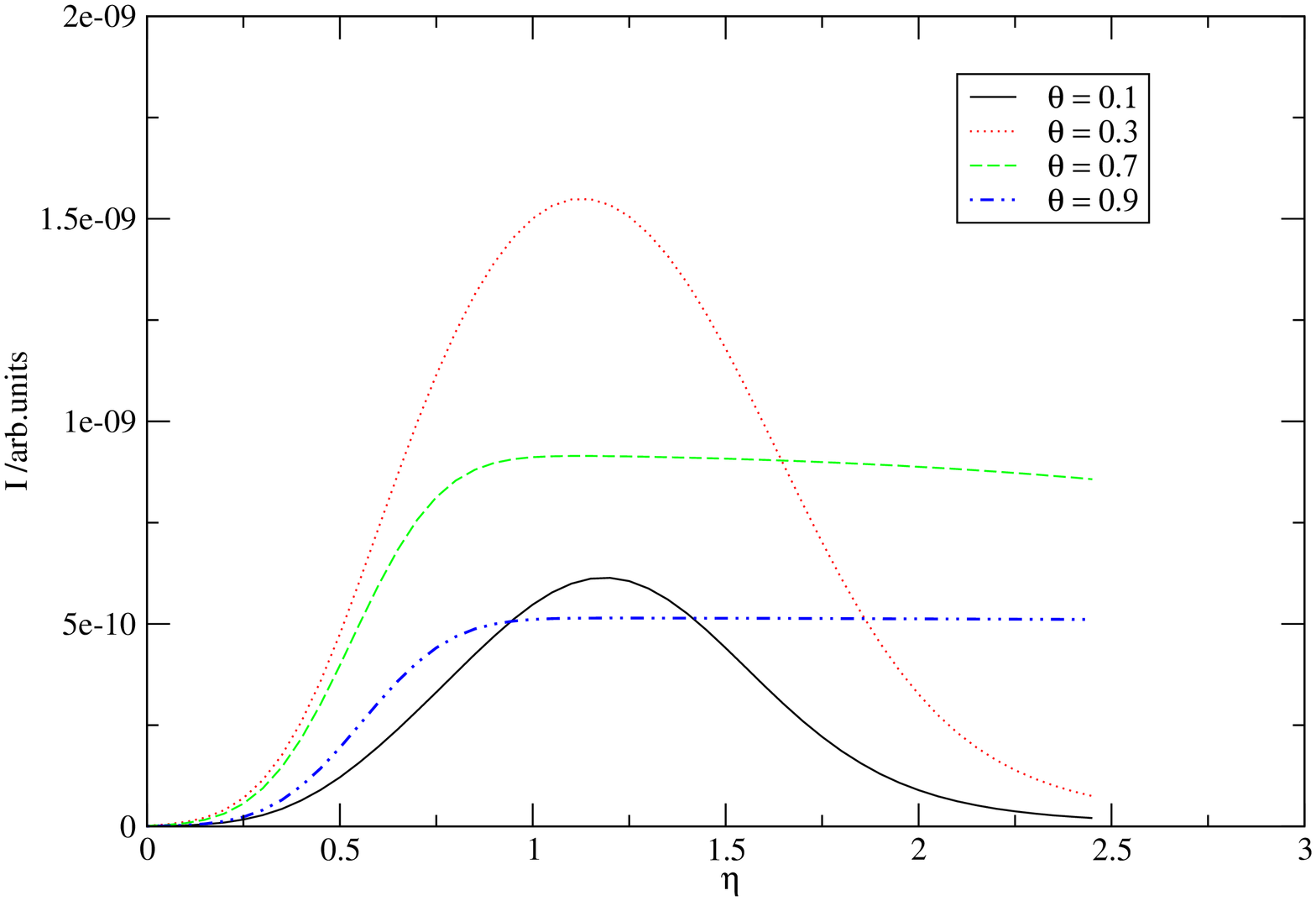}
\caption{anodic current vs $\eta$ for $\alpha$ = 0.3. The values of the various parameters employed (in eV) are as follows: $E^{r} _{r} = 1.0, E^{r} _{ar}(0) = 0.25, E^{r} _{a} (0) = 0.75, \Delta_{||} = 1.5, \Delta_{\perp} = 1.5, \mu = 4.5,\upsilon = 2.0$ }
\label{Ivseta}
 \end{center}
\end{figure}

\begin{figure}[htb]
 \begin{center}
  \includegraphics[scale=0.6,angle=-90]{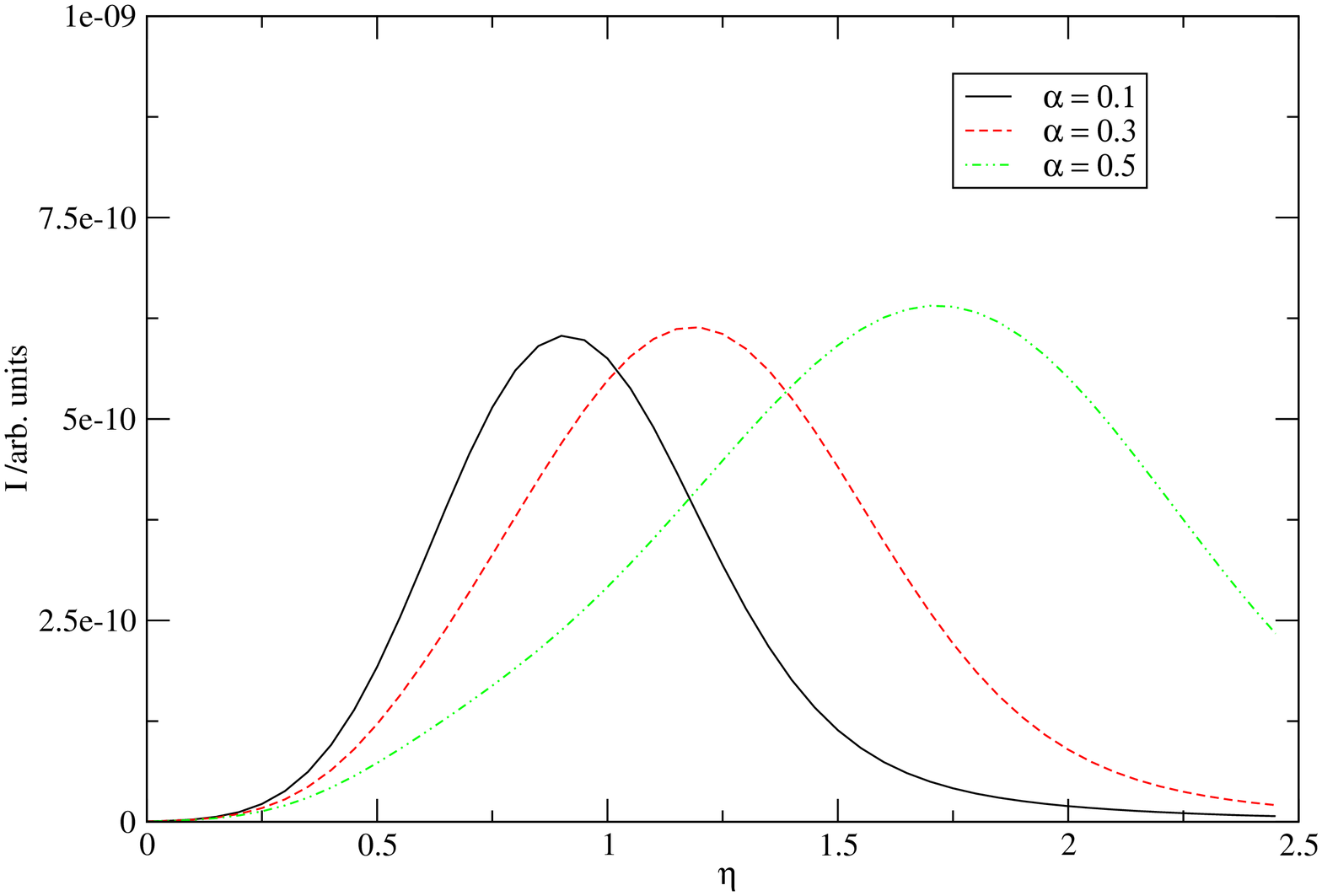}
\caption{ anodic current vs $\eta$ for $\theta$ = 0.1 in the weak coupled regime. The values of parameters (in eV) are as follows: $E^{r} _{r} = 0.6, E^{r} _{ar}(0) = 0.2, E^{r} _{a} = 0.4$ and $v$ = 0.5 eV. }
\label{Ivt1} 
\end{center}
\end{figure}

\begin{figure}[htb]
 \begin{center}
  \includegraphics[scale=0.6,angle=-90]{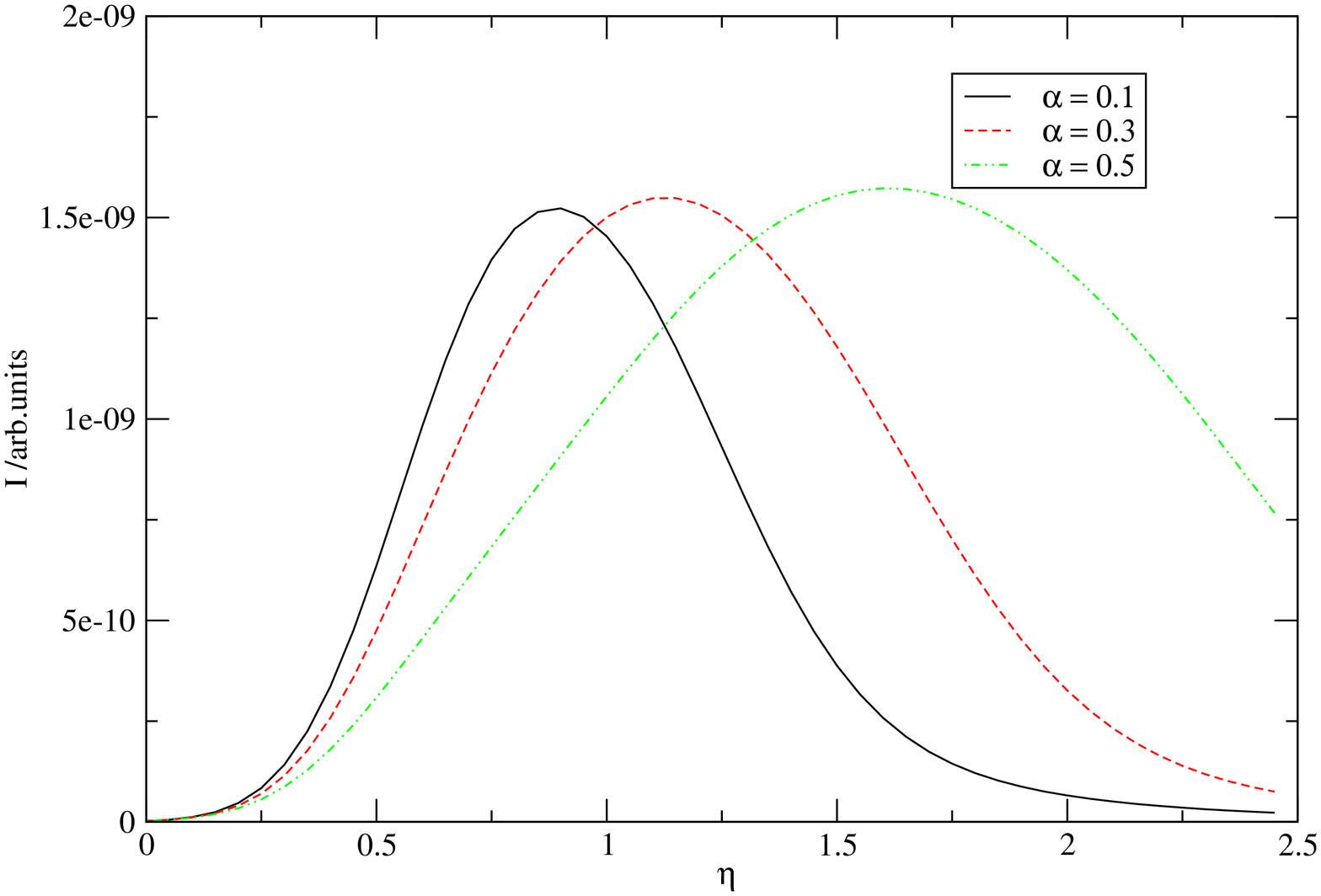}
\caption{ anodic current vs $\eta$ for $\theta$ = 0.3 in weak coupled regime. The values of parameters (in eV) are as follows: $E^{r} _{r} = 0.6, E^{r} _{ar}(0) = 0.2, E^{r} _{a} = 0.4$ and $v$ = 0.5 eV. }
\label{Ivt3}
 \end{center}
\end{figure}

\begin{figure}[htb]
 \begin{center}
  \includegraphics[scale=0.6,angle=-90]{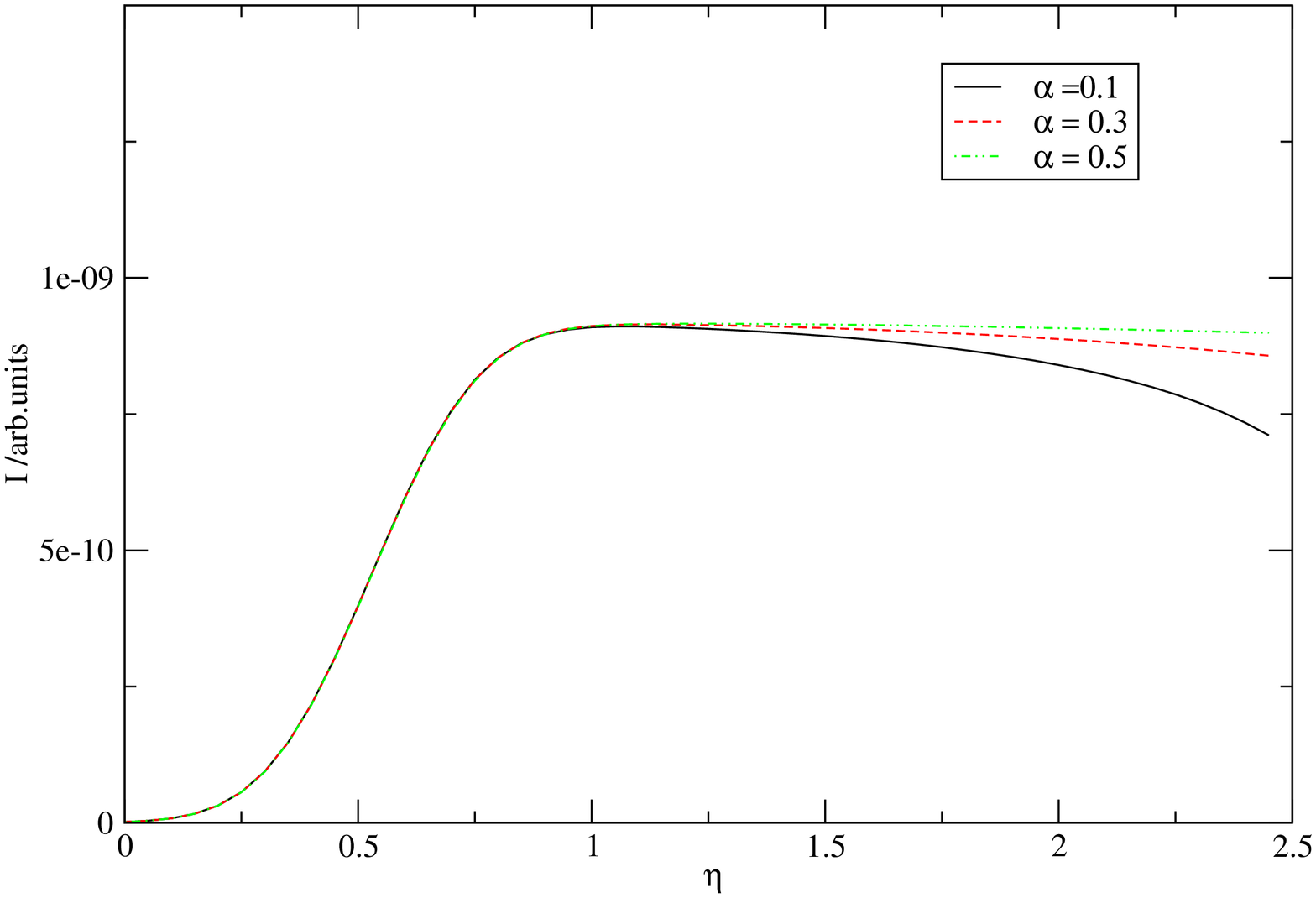}
\caption{ anodic current vs $\eta$ for $\theta$ = 0.7. The values of parameters (in eV) are as follows: $E^{r} _{r} = 0.6, E^{r} _{ar}(0) = 0.2, E^{r} _{a} = 0.4$ and $v$ = 0.5 eV. }
\label{Ivt7}
 \end{center}
\end{figure}

\begin{figure}[htb]
 \begin{center}
\includegraphics[scale=0.6,angle=-90]{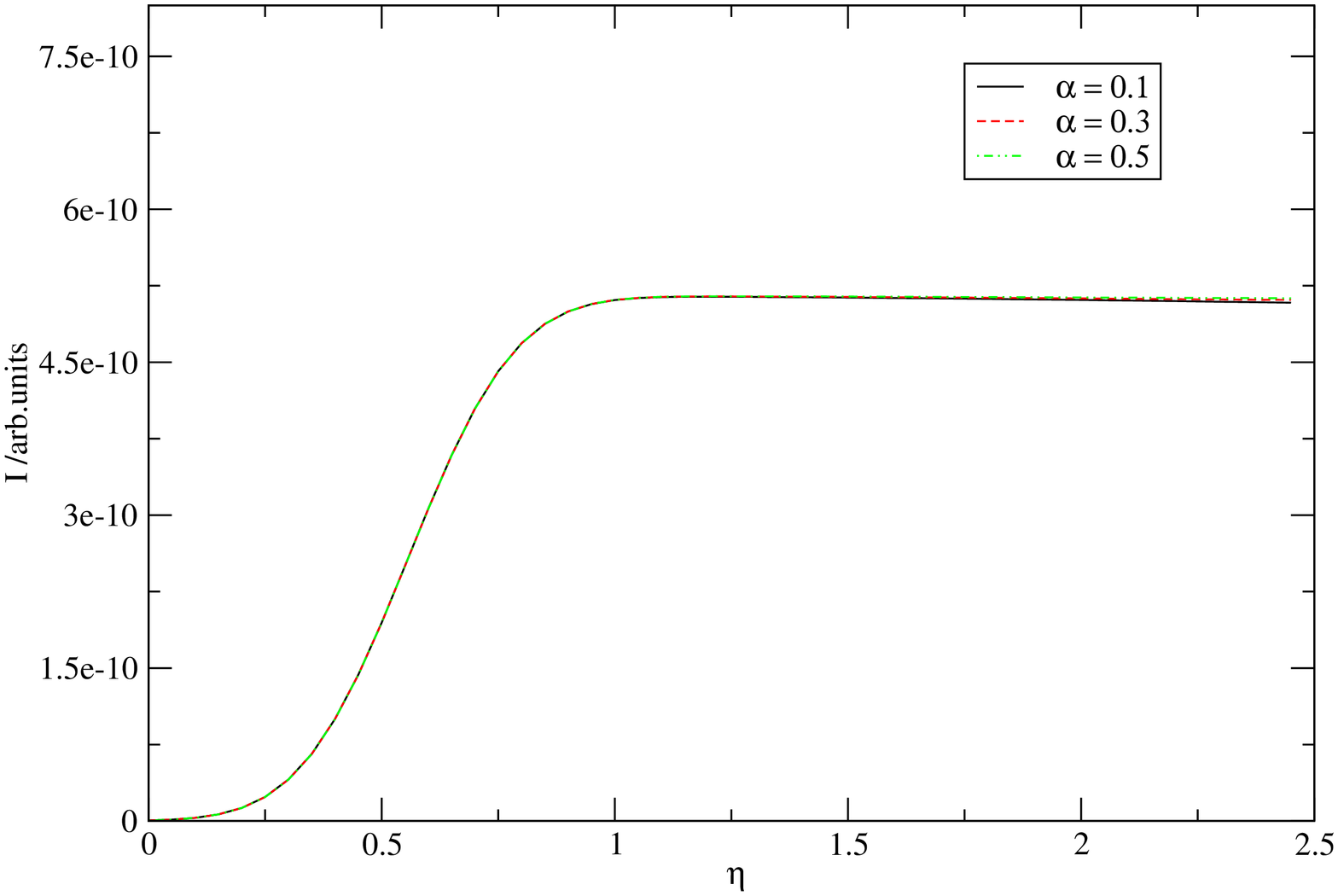}
  \caption{anodic current vs $\eta$ for $\theta$ = 0.9. The values of parameters (in eV) are as follows: $E^{r} _{r} = 0.6, E^{r} _{ar}(0) = 0.2, E^{r} _{a} = 0.4$ and $v$ = 0.5 eV. }
\label{Ivt9}
 \end{center}
\end{figure}

\begin{figure}[htb]
 \begin{center}
\includegraphics[scale=0.6,angle=-90]{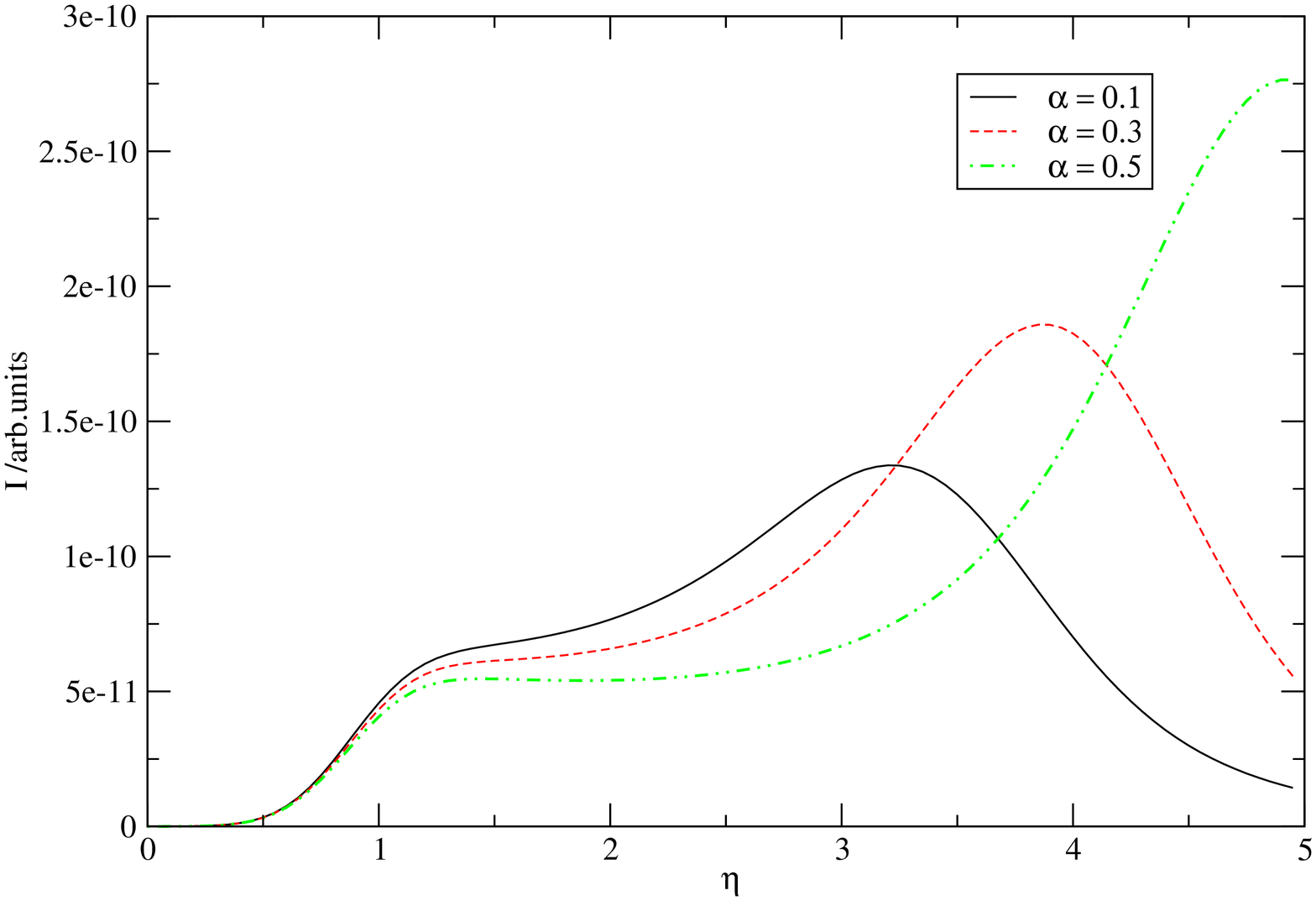}
\caption{anodic current vs $\eta$ for $\theta = 0.1$. The values of the various parameters employed (in eV) are as follows: $E^{r} _{r} = 1.0, E^{r} _{ar}(0) = 0.25, E^{r} _{a} (0) = 0.75, \Delta_{||} = 1.5, \Delta_{\perp} = 1.5, \mu = 4.5,\upsilon = 2.0$}
\label{Iact1} 
\end{center}
\end{figure}

\begin{figure}[htb]
 \begin{center}
  \includegraphics[scale=0.6,angle=-90]{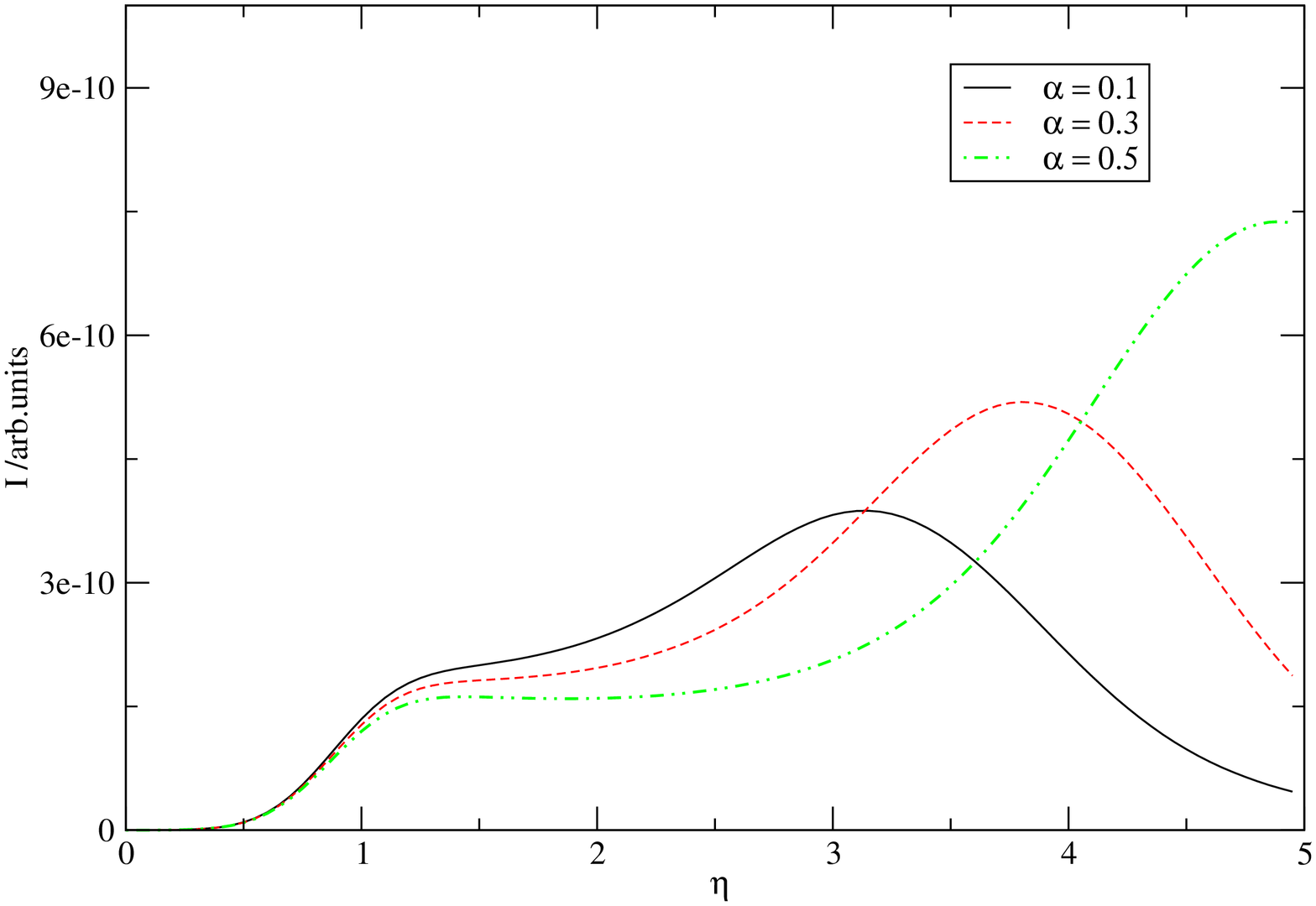}
\caption{anodic current vs $\eta$ for $\theta = 0.3$.The values of the various parameters employed  (in eV) are as follows: $E^{r} _{r} = 1.0, E^{r} _{ar}(0) = 0.25, E^{r} _{a} (0) = 0.75, \Delta_{||} = 1.5, \Delta_{\perp} = 1.5, \mu = 4.5,\upsilon = 2.0$ }
\label{Iact3} 
\end{center}
\end{figure}

\begin{figure}[htb]
 \begin{center}
  \includegraphics[scale=0.6,angle=-90]{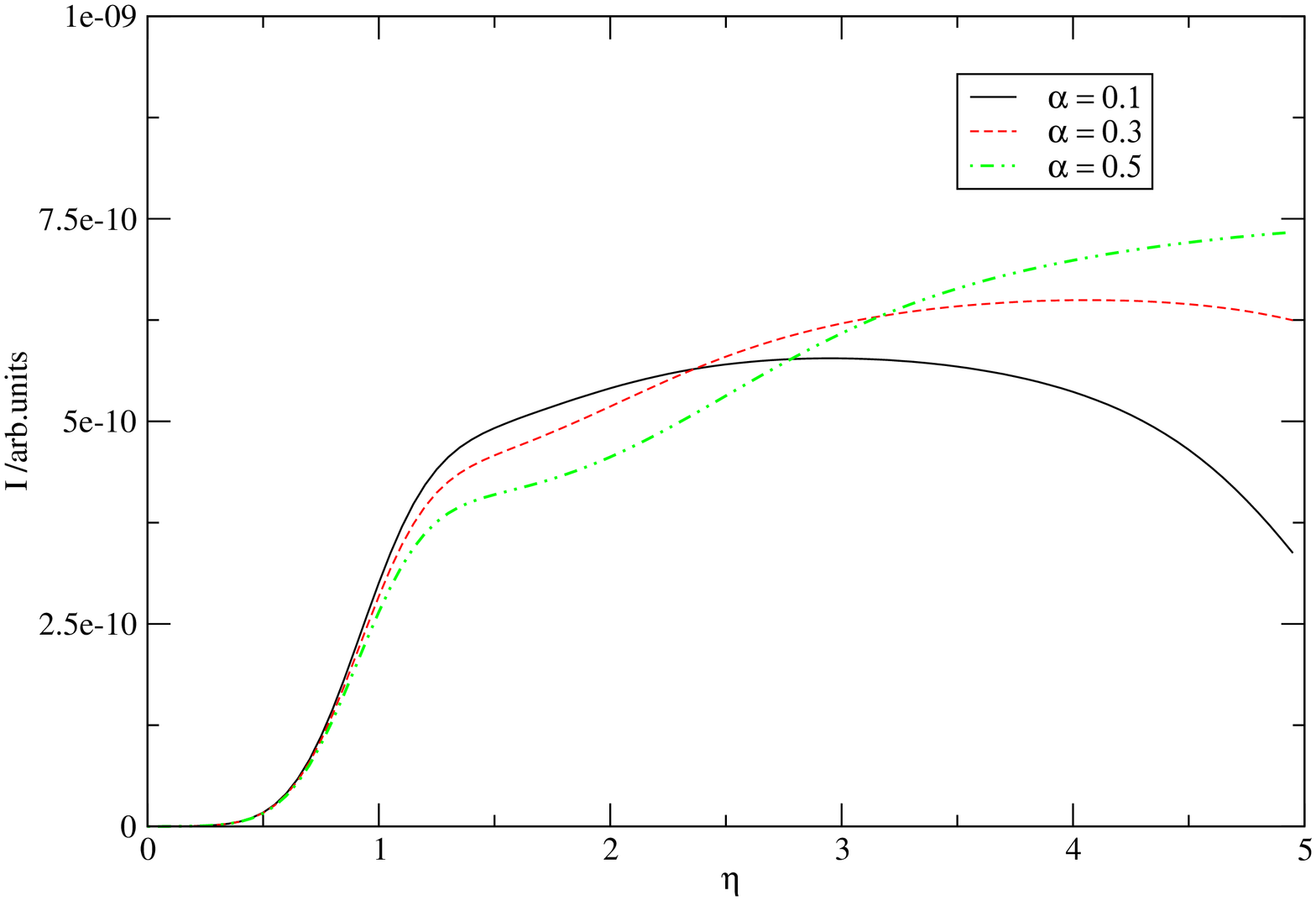}
\caption{anodic current vs $\eta$ for $\theta = 0.7$. The values of the various parameters employed (in eV) are as follows: $E^{r} _{r} = 1.0, E^{r} _{ar}(0) = 0.25, E^{r} _{a} (0) = 0.75, \Delta_{||} = 1.5, \Delta_{\perp} = 1.5, \mu = 4.5,\upsilon = 2.0$ }
\label{Iact7} 
\end{center}
\end{figure}

\begin{figure}[htb]
 \begin{center}
  \includegraphics[scale=0.6,angle=-90]{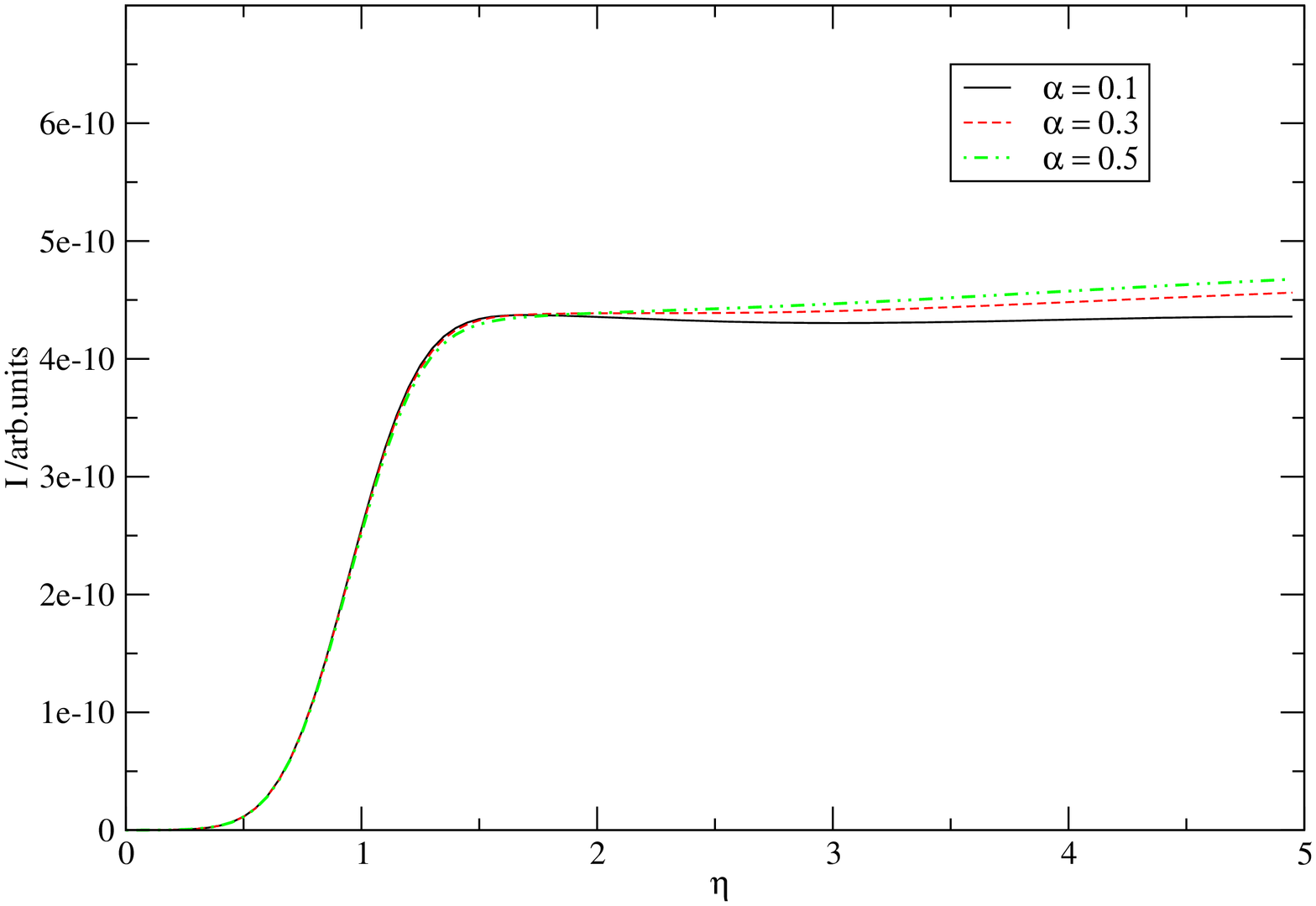}
\caption{anodic current vs $\eta$ for $\theta = 0.9$. The values of the various parameters employed (in eV) are as follows: $E^{r} _{r} = 1.0, E^{r} _{ar}(0) = 0.25, E^{r} _{a} (0) = 0.75, \Delta_{||} = 1.5, \Delta_{\perp} = 1.5, \mu = 4.5,\upsilon = 2.0$ }
\label{Iact9} 
\end{center}
\end{figure}

\begin{figure}[htb]
\begin{center}
 \includegraphics[scale=0.7,angle=-90]{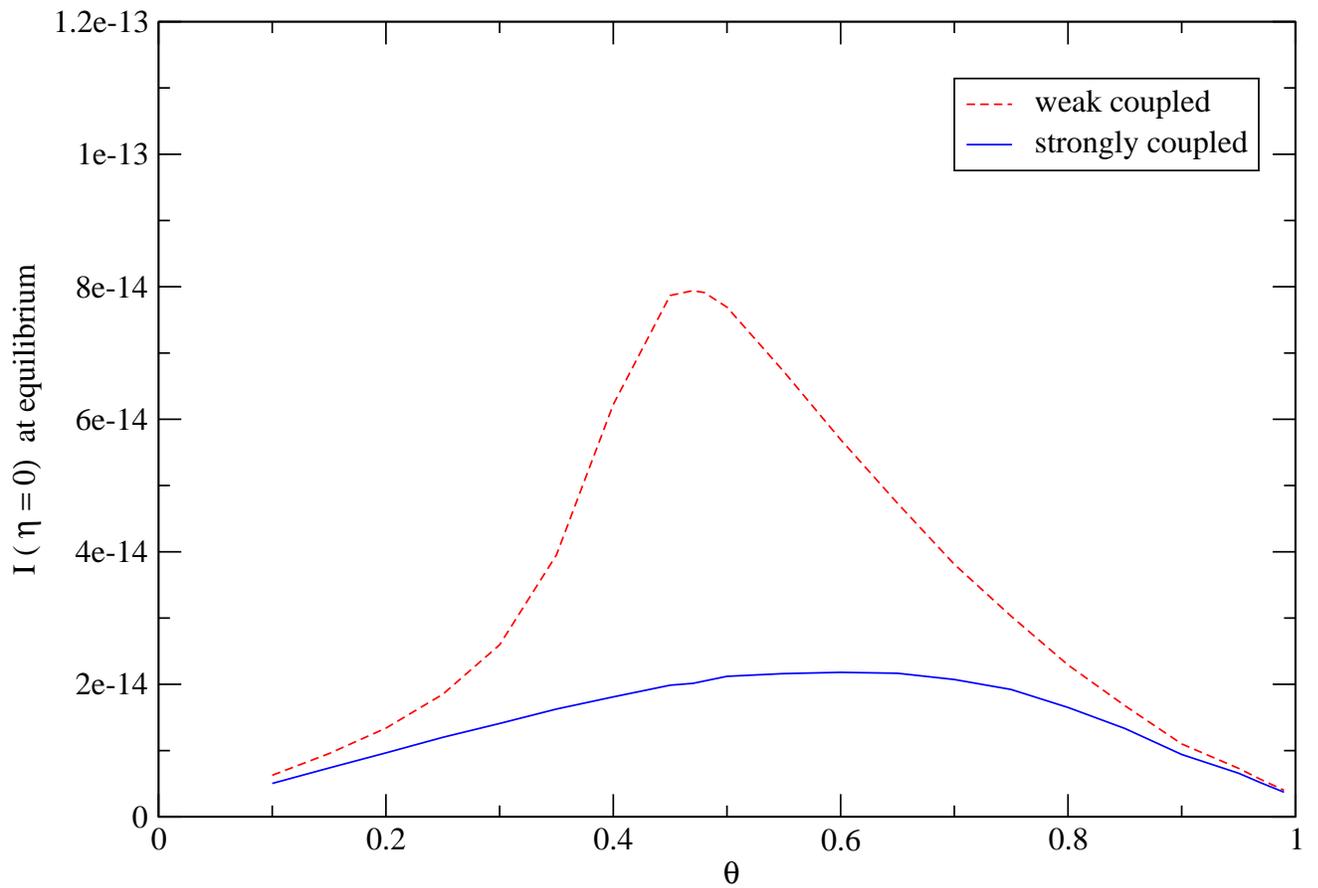}
\caption{Plots showing the equilibrium current at zero overpotential $I_{0}$ vs $\theta$ for strong and weak coupled regime. The values of re-organisation energies were selected be the same for both the curves, $E_{r}$ = 0.6 eV, $E_{a} (0) $ = 0.4 eV, $E_{ar} (0) $ = 0.2 eV }
\label{ci0}
\end{center}
\end{figure}

\begin{figure}[htb]
 \begin{center}
  \includegraphics[scale=0.6,angle=-90]{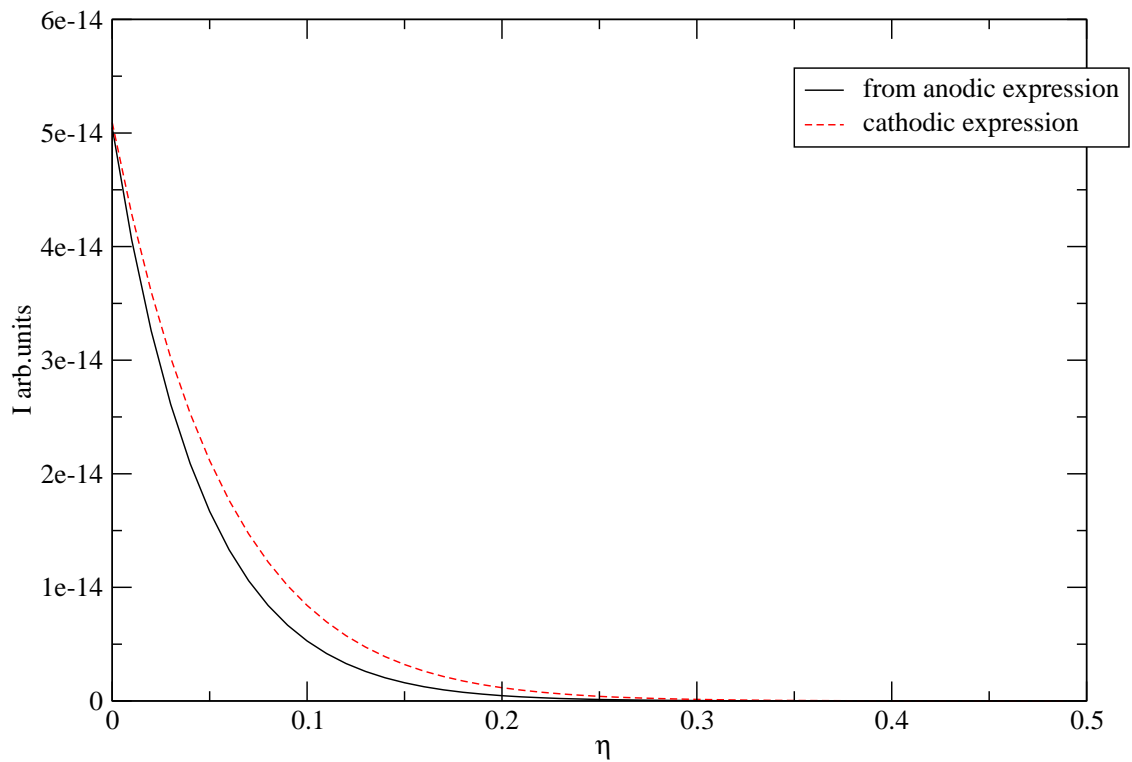}
\caption{Comparison of cathodic current obtained from corresponding expression for cathodic current and from empirical relation of anodic current, $I_{c} = exp[-\beta \eta] I_{a}$. The specific case selected is for $\theta$ = 0.1 and $\alpha$ = 0.5 in the weakly coupled regime. The values of parameters (in eV)  are as follows: $E^{r} _{r} = 0.6, E^{r} _{ar}(0) = 0.2, E^{r} _{a} = 0.4$ and $v$ = 0.5 eV. } 
\label{Iancat} 
\end{center}
\end{figure}

\end{document}